# Novel EDTA-Ligands Containing an Integral Perylene Bisimide (PBI) Core as Optical Reporter Unit

Mario Marcia,[a] Prabhpreet Singh,[b] Frank Hauke,[a] Michele Maggini,[c] and Andreas Hirsch*[a]

[a] *Department of Chemistry and Pharmacy and Institute of Advanced Materials and Processes (ZMP), Friedrich-Alexander University Erlangen – Nürnberg, Henkestrasse 42, 91054 Erlangen (Germany)*
*Fax: (+49)9131-8526864; E-mail: andreas.hirsch@fau.de*
[b] *Department of Chemistry, UGC Centre for Advanced Studies, Guru Nanak Dev University, Amritsar 143005 (India)*
[c] *Department of Chemical Sciences, University of Padua, Via Marzolo 1, 35126 Padua (Italy)*

The synthesis, characterization and metal complexation of a new class of perylene bisimides (PBIs) being an integral part of ethylenediaminetetraacetic acid (EDTA) – is reported. The simplest representative, namely derivative **1a**, was synthesized both by a convergent as well as a direct approach while the elongated derivatives, **1b** and **1c**, were obtained only *via* a convergent synthetic pathway. All these new prototypes of water-soluble perylenes are bolaamphiphiles and were fully characterized by $^{1}$H- and $^{13}$C-NMR spectroscopy, matrix assisted laser desorption ionization – time of flight (MALDI-TOF) mass spectrometry and IR spectroscopy. In order to acquaint for the behaviour in solution of our PBIs bearing dentritic wedges, the simplest derivative, **1a**, was chosen and tested by means of UV/Vis and fluorescence spectroscopy as well as by zeta-potential measurements. A photoexcitation induced intramolecular photo-electron transfer (PET) can be observed in these molecules. Therefore potential applications as sensor can be imagined. Model compound **1a** efficiently coordinates trivalent metal cations both in water and in dimethyl sulfoxide (DMSO). Significantly, the effects of the complexation strongly depend on the aggregation state of the PBI molecules in solution. As a matter of fact, in water, the presence of $M^{3+}$ ions triggers the formation of light emitting supramolecular aggregates (excimers). On the other hand, in DMSO-rich solutions metal complexation leads to the suppression of the PET and leads to a strong fluorescence enhancement.

## Introduction

Perylene bisimides (PBIs) represent a class of aromatic chromophores with a series of interesting properties. From an industrial point of view they find wide spread applications as high-performance pigments[1] due to their fascinating optoelectronic and electrochemical properties.[2] Therefore, they represent integral components in photovoltaic devices,[3] sensors,[4] OLEDs,[5] OFETs[6] as well as dye-lasers[7] and have been used as building blocks for liquid crystals,[8, 9] for membrane labelling[10] and in the photodynamic therapy.[11]

PBIs are usually insoluble in water. Therefore, several attempts have been performed in order to increase the solubility in aqueous media, either with functionalization in the bay region[12-16] or at the side imide groups.[17-25] Recently, we have introduced a series of highly water-soluble PBIs containing anionic (**R$^1$**) or cationic ($\mathbf{R}^2$) Newkome dendronized substituents – representative examples are given in Figure 1. Their aggregation behavior in dependence of pH value, ionic strength and concentration[26-28] has been investigated and based on their pronounced solubility, their ability to exfoliate and stabilize single walled carbon nanotubes (SWCNTs)[29] and graphene in water[30, 31] has been demonstrated.

Here, we report the synthesis, characterization and properties of a new family of PBI-based amphiphiles. These derivatives belong to a novel class of PBI–based surfactants, where di- and polyamine spacers have been used as building blocks for the construction of arrays of oligocarboxylic acid molecules. They can be regarded as elongated ethylenediaminetetraacetic acid (EDTA) ligands containing an integral aromatic perylene bisimide (PBI) core as optical reporter unit.

**Fig.1** Structure of 1$^{st}$ generation Newkome dendronized PBIs

**Fig. 2** Structure of EDTA-PBIs, **1a – c**.

In addition, these EDTA-PBIs (Figure 2) are characterized by a very small periphery in comparison to the Newkome dendronized ones (Figure 1). The presence of the big bulky substituents at both termini of the PBI is expected to strongly determine their aggregation behavior in solution, leading to the formation of monomers even at high concentrations. Although this feature is generally desirable, it could be a hurdle in certain specific applications, such as the dispersion/exfoliation of carbon allotropes where strong aggregation of the exfoliating agent is generally pursued[32] in order to optimise the production of individualized graphene layers/carbon nanotubes in solution. Therefore, smaller side groups emphasize the importance of the molecular core, still ensuring good water solubility and offering supplementary chelation properties. Moreover, the terminal units of PBIs **1a** – **c** provide an additional feature as they enable a facile metal-chelation capacity in both aqueous and organic solutions. Due to the presence of the PBI core as optical reporter unit, the complexation of various metal cations can be studied easily by means of optical analytical techniques, such as UV/Vis and fluorescence spectroscopy.

## Results and Discussion

For the synthesis of **1a**, two different chemical strategies were pursued. The first route (Scheme 1, top) is based on a classical convergent synthetic approach, where the selectively functionalized – protected amine **2** was condensed with perylene-3,4,9,10-tetracarboxylic dianhydride (PTCDA), yielding the *tert*-butyl protected version **3**, which was subsequently deprotected in trifluoroacetic acid (TFA) – for further details regarding the synthesis of compound **2** see electronic supporting information, ESI. The free acid **1a** is obtained with an overall yield of 21.6 % (based on the starting material putrescine). In addition, this approach is quite time consuming and employs several selective protection and deprotection steps, which impede a straightforward scale-up of the synthesis.

The second strategy (Scheme 1, bottom) involves the direct alkylation of a PBI-based precursor amine and is performed in only three steps: a) the condensation of PTCDA with putrescine, which yield the perylene bisamine **4**; b) the alkylation of **4** to the corresponding tetraester **3** and c) the quantitative removal of the *tert*-butyl protection group releasing the free acid **1a** with an overall yield of 11.6 %.

The direct synthesis is primarily based on low cost chemicals and does not require the use of protected amine precursors such as amine **2**. Therefore, the outlined synthetic route provides the basis for a cost efficient large scale synthesis of these novel EDTA-PBI derivatives.

Nevertheless, the drawback of this approach is the alkylation of the derivative **4** due to its low solubility in the common suitable solvents for $S_N2$ reactions (*e.g.* $CH_3CN$), which is responsible for the relatively moderate overall yield.

With this knowledge at hand, the higher branched EDTA-PBIs **1b** and **1c** where synthesized solely according to a convergent approach (Scheme 2 and 3). The synthesis of the respective precursor amines **2**, **11** and **17** is reported in the electronic supporting information. All the derivatives were fully characterized by $^{1}$H- and $^{13}$C-NMR and IR spectroscopy as well as by mass spectrometry and elemental analysis.

For the detailed investigation of the aggregation behavior of these novel compounds and for the direct comparison of their results with other commonly available chelating agents, the simplest derivative **1a** has been chosen and solution based optical measurements were carried out.

Compound **1a** exhibits a good solubility in basic (NaOH$_{(aq)}$, $c$ = $1\cdot10^{-3}$ M) as well as acidic (HCl$_{(aq)}$, $c$ = $1\cdot10^{-3}$ M) aqueous media. Among the common organic solvents, **1a** is very good soluble only in dimethyl sulfoxide (DMSO). The solubility in organic solvents (such as DMSO; *N*-methyl pyrrolidone, methanol; *N*,*N*-dimethyl formamide and diethyl ether) can be drastically increased by the addition of acids, like TFA, formic acid or hydrochloric acid (HCl$_{(aq)}$).

In the latter case, though, the introduction of water leads to a strong aggregation of the EDTA-PBI derivative by intermolecular $\pi$–$\pi$ stacking interactions. These induced aggregation phenomena were investigated by $^{1}$H-NMR spectroscopy (electronic supporting information) as well as absorption and emission spectroscopy.

Two aqueous (diluted acid and basic medium) and two organic (DMSO and DMSO with addition of TFA) solvents have been chosen to study the aggregation behavior of EDTA-PBI derivative **1a**. For each system the most relevant optical properties have been determined and are collected in Table 1.

**Table 1** Optical properties of **1a** in aqueous and organic solvents

| **Property** | **$NaOH_{(aq)}$** | **$HCl_{(aq)}$** | **DMSO** | **$H^+$-DMSO**[d] |
|---|---|---|---|---|
| **log $\varepsilon_{max}$ ($M^{-1}\cdot cm^{-1}$)** | 4.38[b] | 4.25[b] | 4.53[c] | 4.76[c] |
| **$\lambda_{abs, max}$ (nm)** | 543, 500 | 549, 481 | 528,494,460 | 528,494,460 |
| **$\lambda_{fluo, max}$ (nm)** | 587,546 | 587, 546 | 580, 543 | 580, 543 |
| **Φ (%)**[a] | $2.3 \pm 0.2$ | $0.27 \pm 0.05$ | $9.1 \pm 1.1$ | $63.9 \pm 10.6$ |

[a]*The fluorescence quantum yield (Φ) is calculated taking as reference Fluorescein in NaOH 0.1M*[33]. [b]*ε calculated at 500 nm.* [c]*ε calculated at 494 nm.* [d]*DMSO with addition of TFA, $3\cdot10^{-4}$ M*

Normally, PBIs without functional moieties attached to the aromatic aromatic core can exist as aggregated as well as monomeric species in solution, depending on the solvent used.

**Scheme 1** Convergent (top) and direct (bottom) synthesis of EDTA-PBI **1a**. i) Imidazole, $Zn(OAc)_2$, 110 °C, 4 h, yield: 64 %; ii) TFA:$CH_2Cl_2$ (2:1), rt, 5 days, yield: 84 %; iii) toluene, reflux, 4 h, yield: 89 %; iv) acetonitrile, DIPEA, *tert*-butyl bromoacetate, 60 °C, 24 h, yield: 13 %; v) formic acid, RT, 2 days, yield: 100 %.

**Scheme 2** Convergent synthesis of EDTA-PBI **1b**. i) Imidazole, $Zn(OAc)_2$, 110 °C, 4 h, yield: 53%; ii) TFA:$CH_2Cl_2$ (2:1), RT, 5 days, yield: 82%

**Scheme 3** Convergent synthesis of EDTA-PBI **1c**. i) Imidazole, $Zn(OAc)_2$, 110 °C, 4 h, yield: 56%; ii) TFA:$CH_2Cl_2$ (2:1), RT, 5 days, yield 74 %

In the case of monomeric PBIs, the distinct absorption band relative to the electronic transition $S_0 \rightarrow S_1$ is generally found in the range 400 – 600 nm and is characterized by three well resolved vibronic peaks. The predominant two features, located at ≈ 530 nm (0,0) and ≈ 490 nm (0,1), respectively, exhibit a ratio greater than 1.6.[34] Upon aggregation, pronounced modifications appear in the spectra. The intensity of the (0,0) peak is reduced, while that of the (0,1) peak increases accompanied by a pronounced reduction of the absorption coefficient.[17] The ratio between the (0,0) and (0,1) peaks reaches a value of ≈ 0.7 or lower in the case of strong aggregation.[29, 35] As a result, the colour of the solution changes from orange to red.

As depicted in Figure 3, **1a** is significantly aggregated, at room temperatures, both in diluted basic ($NaOH_{(aq)}$, $c$ = $1 \cdot 10^{-3}$ M) and acid ($HCl_{(aq)}$, $c$ = $1 \cdot 10^{-3}$ M) aqueous conditions, due to the strong $\pi$–$\pi$ stacking interactions of the perylene aromatic cores.

**Fig. 3** Absorption and fluorescence profiles of **1a** in $NaOH_{(aq)}$ (red curve) and in $HCl_{(aq)}$ (black curve), $c = 5 \cdot 10^{-6}$ M.

The behavior of **1a** can be explained as follows. In basic aqueous conditions, pH ≈ 11, the EDTA-PBI is completely deprotonated (EDTA is completely deprotonated at pH > 10.6). Therefore, the aggregation proceeds most certainly *via* the formation of PBI stacked systems where each molecule is rotated with respect to its nearest neighbors. In such a way, the electrostatic repulsion of the residual charges situated at the periphery of each molecule would be minimized.[36]

In acid aqueous solution the aggregation is even more pronounced. As a matter of fact, at the pH under investigation (pH ≈ 3) it is impossible to have a complete protonation of **1a** due to the fact that the isoelectric point of **1a** is located at a value of pI = 2 or below.[37] We expect thus that **1a** exists in an equilibrium between several partly protonated species ($H_xEDTA^{y+}$, where $x < 4$ and $y < 2$) in solution. Therefore, no sufficient electrostatic repulsion is believed to hinder the aggregation by $\pi - \pi$ stacking successfully. Moreover, the absorption spectrum of **1a** in diluted $HCl_{(aq)}$ shows a great broadening of the (0,1) peak in comparison to solutions of **1a** in diluted $NaOH_{(aq)}$ at the same concentration of the surfactant. The enlargement of the (0,1) peak accounts generally for an extended aggregation in solution[17] and this confirms consequently that in diluted $HCl_{(aq)}$ the aggregation takes place easier. In addition, further aggregation might derive from intermolecular hydrogen bonds between the protonated carboxylic moieties.

Furthermore, H-aggregates[38] are formed in aqueous solution both in diluted $NaOH_{(aq)}$ and $HCl_{(aq)}$. In both cases, the presence of PBI-dimers[39] is excluded.

In order to get more insight into these aggregation phenomena, we decided to perform additional Zeta potential measurements. The results are collected in Table 2.

Zeta potential (ζ) measurements are used in order to predict the stability of dispersions. According to Greenwood *et al.*[40], a Zeta potential value higher than ± 30 mV indicates that a colloidal suspension is rather stable; moderate instability is characterized by a ζ value comprised between ±10 mV and ± 30 mV while extensive coagulation/flocculation occurs for ζ value around 0 mV.

**Table 2** Zeta potential measurements of **1a** in aqueous conditions ($c = 1 \cdot 10^{-6}$ M)

| Property | $NaOH_{(aq)}$, $c = 1 \cdot 10^{-3}$ M | $HCl_{(aq)}$, $c = 1 \cdot 10^{-3}$ M |
|---|---|---|
| Zeta potential (mV) | -13.3 ± 2.6 | 0.7 ± 0.7 |

From the results collected in Table 2, it is possible to conclude that **1a** is more stable in diluted $NaOH_{(aq)}$ compared to diluted $HCl_{(aq)}$. However, in both cases the suspensions are rather unstable ($\zeta < 15$ mV).

The values of the UV and Zeta potential measurements are also corroborated by the fluorescence quantum yield data. Emission spectroscopy provides further insights into the equilibrium between monomeric and aggregated species. The dominant fluorescence can always be traced back to the monomeric species. In this case the emission profile is also the mirror image of the absorption spectrum. As aggregation takes place (Figure 3) the absorption spectrum changes and therefore the emission does not appear any longer as mirror image of the absorption spectrum.

As evinced from Table 1, the fluorescence quantum yield (Φ) is approximately ten times lower for **1a** in diluted $HCl_{(aq)}$ compared to **1a** in diluted $NaOH_{(aq)}$. Such drastically decreased Φ values in diluted $HCl_{(aq)}$ solutions might, however, appear counterintuitive. As a matter of fact, the tertiary amine functionalities, situated at both termini of the PBI acceptor aromatic core, determine a pronounced fluorescence quenching as a result of an intra-molecular photo-induced electron transfer (PET).

Similar PBI dyes, which also bear bisamine functionalities linked to the perylene bisimide aromatic core, are generally characterized by an increase of the fluorescence either upon acid addition[19] or by chemical derivatization[41] and also in the presence of metal cations.[42] However, in diluted $HCl_{(aq)}$ it is observed that the aggregation process prevails over the amine protonation. The aggregation by π–π stacking contributes therefore to an overall fluorescence quenching even if the PET process might be partly suppressed.

When the aggregation in solution is prevented it is possible to examine the PET process in detail and to understand how it influences the fluorescence of PBI **1a**. For this study, we decided to investigate a solution of **1a** in DMSO, in the presence or absence of an acid. TFA was chosen due to the fact that the addition of diluted $HCl_{(aq)}$ would add water to the system, which leads to a more pronounced aggregation of **1a**.

First of all, in pure DMSO the behavior of **1a** is quite different (Figure 4) compared to that in aqueous solutions at room temperature. As a matter of fact, the (0,0) peak is higher in intensity with respect to the (0,1) peak and their relative ratio is about 1.2. This value is lower than the threshold for completely monomeric PBIs (0,0)/(0,1) ratio ≈ 1.6 and therefore we conclude that **1a** is present as a mixture of the monomeric and aggregated species in solution.

Due to the residual aggregation and because of the PET process, the fluorescence quantum yield of **1a** in DMSO remains pretty low (Φ < 10 %) but five times higher with respect to the value measured in $NaOH_{(aq)}$ solutions, where the EDTA-PBI is strongly aggregated. Upon addition of a small amount of TFA to a DMSO solution of **1a** the fluorescence increases (Figure 4) as a result of the protonation of the two tertiary amines, which lowers the energy of their nonbonding orbital below that of the HOMO of the PBI, allowing to switch-on of the fluorescence of the chromophore.[43] In such conditions the fluorescence quantum yield increases drastically (Table 1). Concomitantly, due to protonation of the tertiary amines, the aggregation *via* π–π stacking is lowered and the solubility of **1a** is increased.

**Fig. 4** Absorption and fluorescence profiles of **1a** in DMSO (red curve) and in acid-DMSO (black curve), $c = 5\cdot10^{-6}$ M.

The possibility of switching on/off the fluorescence is remarkable and offers access to a broad range of applications for **1a** and analogues PBIs.[44-47] Moreover, as described elsewhere,[48, 49] it might be possible to correlate the fluorescence intensity to the pH and therefore obtain information about the acid/base equilibrium in solution. In our case, though, after the strong aggregation of **1a** took place, titration with either $NaOH_{(aq)}$ or $HCl_{(aq)}$ led to very little modifications of the fluorescence intensity and therefore hindered a reliable determination of the six $pK_a$s values of **1a**. However, due to the presence of a ligand periphery attached to a PBI reporter unit, derivative **1a** allows the possibility to investigate the coordination of metal ions in solution by means of spectroscopic measurements.

Both naphthalene[50, 51] and perylene[52-58] based chelating agents have been reported, but to the best of our knowledge none of them with an elongated EDTA-like structure containing an integral PBI unit as optical receptor unit.

As a matter of fact, the tailor-made periphery of **1a** was designed to efficiently chelate metal cations. The lock-and-key interaction of the metal cations with the tertiary amine functionalities at both sides of the EDTA-PBI will result in an energy/electron transfer, which will be transmitted to the central PBI core *via* the PET process. By monitoring changes in the fluorescence of **1a** it can be defined a set of parameters in order to determine the influence of different metals on the EDTA-PBI chromophore.

Of course, the aggregation by π–π stacking of **1a** in water and organic solvents will strongly influence the complexation of metal cations in solution. Such hydrophobic interactions could be responsible for the formation of a specific enviroment in solution which allows the easier coordination of certain metal cations and not those expected for similar chelating agents, such as EDTA or iminodiacetic acid, IDA (Figure 5).

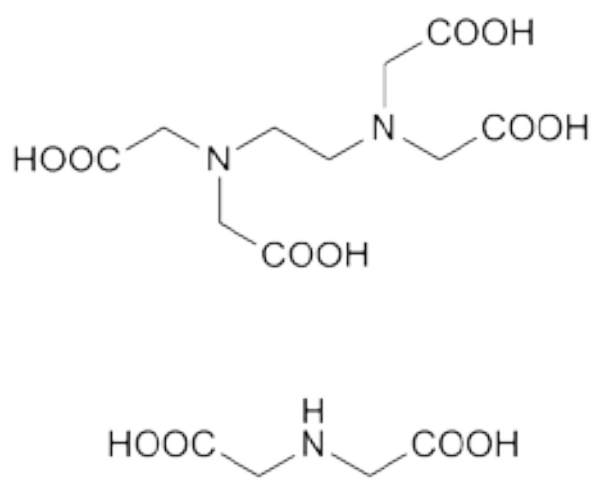

**Fig. 5** Structures of EDTA (top) and IDA (bottom).

The structure of **1a** resembles that of EDTA with the insertion of an integral PBI core between the two chelating moieties. Upon insertion, the separation between the two tertiary amine substituents is enhanced and an intramolecular EDTA-like coordination geometry (hexadentate ligand) is lost. However, the octahedral chelating geometry may be ripristinated by intermolecolar interactions in solution, *e. g.* due to aggregation of the PBI molecules. Nevertheless, if in the aggregates the peripheral groups remain remotely enough, the coordination geometry of **1a** should resemble more that of IDA (tridentate ligand).

Due to this complexity, we assume that the coordination geometry of **1a** in solution would be neither only that of IDA or EDTA, but more reasonably a mixture of both. The complexation of metal cations was investigated both in aqueous and organic media. Working in water based systems, on the one hand, is generally preferred for biological applications although in our case this means a stronger aggregation of the PBI aromatic moieties. In organic solvents, on the other hand, the π–π stacking is generally hindered but only niche applications can be addressed. Moreover, in non-aqueous systems it is more difficult to account for acid/base equilibria which may affect the complexation.

In order to gain the most comprehensive view of the affinity of **1a** for metal cations in aqueous solutions at room temperature, thirty common salts have been employed (electronic supporting information).The first study was accomplished in distilled water, where **1a** is soluble only at concentrations below $10^{-5}$ M. As mentioned before, in aqueous conditions aggregation takes place easily. This has the following drawbacks. Above all, the fluorescence of **1a** is greatly quenched and therefore it is more difficult to discern the origin of the PET process. Moreover, the introduction of ion species in a solution where **1a** is already aggregated might also favor the formation of extended aggregates, which could be defined as supra-molecular polymers in solution.[59-67]

In a typical experiment, to a solution of **1a** in distilled water ($c = 5 \cdot 10^{-6}$ M), 10 equivalents of the chosen metal cations have been added at room temperature. The solution was agitated mechanically for five minutes by means of a platform shaker and then the fluorescence spectrum was measured and compared to a reference solution of **1a** in distilled water. The integrated fluorescent intensity in presence (F) and absence ($F_0$) of the metal cation were determined subsequently and the results were compared in terms of their ratio ($F/F_0$). A value of $F/F_0$ higher than 1 indicates an enhancement of the fluorescence of the PBI (EF effect), while a value lower than 1 indicates a fluorescence quenching (QF effect).

The influence of the metal counter-ions has been also tested by addition of 10 equivalents of different sodium salts to a standard solution of **1a** in distilled water. The results show that different anions contribute on average to the same EF/QF in presence of the same metal cations and therefore a counter-ion effect is generally excluded. The results of the complexation experiments for **1a** in aqueous environment have been divided according to the charge of the respective cations (see supporting information). First of all, it can be noticed that the addition of alkaline metal cations leads to an enhancement of the fluorescent of **1a** in water solutions only in presence of $K^+$ and $Cs^+$, while this does not happen for $Li^+$ and $Na^+$. Earth-alkaline metal ions have a negligible effect or induce a little decrease in the fluorescence intensity of **1a**.

These preliminary results suggest that most of the alkaline and earth alkaline cations are too small to interact efficiently with the chelating part of the EDTA-PBI. This might be explained considering the ionic radius of alkaline and earth alkaline metal cations. Actually $K^+$ and $Cs^+$ possess the highest values. Therefore, it might be argued that in the particular environment created by **1a**'s micelles in water solution only big cations can efficiently bind the nitrogen atom of the ligand side groups and slightly affect the PET process. The addition of transition metals and lanthanides promotes, instead, a drastic fluorescence quenching ($F/F_0 < 0.16$), with the exception of $Au^{3+}$ ($F/F_0 \approx 0.3$) and $Ag^+$ ($F/F_0 \approx 0.7$). In particular the most pronounced QF values were obtained for addition of copper ions (both $Cu^+$ and $Cu^{2+}$, QF = 0.0066) and trivalent cations ($M^{3+}$ QF $\approx$ 0.01). Such a drastic fluorescent quenching is to be attributed to electron/ energy transfers between the *d* electrons of the transition metals and the EDTA-PBI accepting molecules. As a matter of fact, the presence of transition metals or lanthanide cations produces localized redox reactions (electron transfer) between the metal centers and the PBI moieties occur which result in the dramatic quenching of the fluorescence of **1a**. Additionally, the metal cations might bridge vicinal PBI-micelles causing the formation of extended aggregates. This highly impressive fluorescence quenching is observed mainly for bivalent metal cations with the exception of $Pb^{2+}$, whose QF is not as pronounced ($F/F_0 \approx 0.4$) and also for $Cu^+$. Actually, $Pb^{2+}$ is not a transition metal and a lower influence on the emission of the EDTA-PBI should be expected. Moreover, it is important to underline that a remarkable quenching of the fluorescence of **1a** can also be observed in the presence of metals with a $d^{10}$ electronic configuration which cannot profit from ligand field effects, such as $Zn^{2+}$, $Cd^{2+}$ and $Hg^{2+}$. Such metals do have a complete filled *d* level and cannot take part in any direct electron transfer processes. However, it has been reported that depending on the geometry of the complex in solution other types of electron transfer processes may contribute to the quenching of the fluorescence of the chromophore.[68] In addition, De Santis *et al.*[69] showed that $Zn^{2+}$ has a specific affinity towards carboxylic acid groups which may lead to a strong fluorescence quenching of the chromophore.

The addition of trivalent metal cations contributes, as well, to a global quenching of the fluorescence of **1a**. In particular, the QF values are comparable to those recorded for copper cations ($F/F_0 << 0.1$). Nevertheless, the quenching of fluorescence is now accompanied by a strong modification of the emission spectrum of the EDTA-PBI molecule. As a matter of fact, the presence of trivalent metal cations, induced the formation of a new broad band in the emission spectrum of **1a** (Figure 6), which may account for the formation of emitting aggregated species (excimers) in solution.[70-75]

Furthermore, after a decantation of 2-3 days, the solutions of **1a** titrated with 10 equivalents of $M^{3+}$ become transparent and a solid precipitant was formed. The new broadened feature present in the emission spectrum of **1a**, which is not present in the normal emission spectrum of **1a** in aqueous conditions, is located between 660 – 680 nm and the $\lambda_{max}$ of the peak shifts depending on the metal cation added ($Sm^{3+}$ $\lambda_{max} \approx 660$ nm, QF ≈ 0.05; $In^{3+}$, $Gd^{3+}$ and $La^{3+}$ $\lambda_{max} \approx 670$ nm, QF ≈ 0.01; $Fe^{3+}$ and $Sc^{3+}$ $\lambda_{max} \approx 675$ nm, QF < 0.01; $Ce^{3+}$ $\lambda_{max} \approx 680$ nm, QF ≈ 0.01). The formation of this new band in the emission spectrum of PBIs has been extensively discussed in the literature. According to Wang *et al.*[35, 76-78] who examined the self-organization of a PEG functionalized PBI in chloroform solution, such a red shifted band in the emission spectrum of PBIs may be attributed to the fluorescence properties of molecular assemblies of increased size. The same conclusions have also been outlined by Neuteboom and co-workers[79] when studying the optical features of polytetrahydrofuran substituted PBI polymer in ODCB. Arnaud *et al.*,[71] Würthner *et al.*,[70, 80, 81] Datar *et al.*,[82] and Yagai *et al.*[75] reported as well the formation of this red-shifted band due to self-assembled/polymeric PBI in solutions.

**Fig. 6** Fluorescence spectra of **1a** in water after addition of 10 eq. of $M^{3+}$ ($Sm^{3+}$ is not shown for clarity, due to its higher QF value).

To probe the stability of these metal-induced aggregated emitting species in solution, temperature dependent fluorescence measurements have been performed. As depicted in Figure 7, an increase in temperature (from 25 °C to 75 °C), leads normally to an increase of the $F/F_0$ value and this correlates with a lower aggregation in solution. This observation is corroborated by a decrease in intensity and eventually in the disappearance of the band situated between 660 – 680 nm in the fluorescence spectrum. The only exception to this pattern is found for $In^{3+}$.

For this cation in fact, the $F/F_0$ ratio remains practically constant even at higher temperatures and the band due to PBI excimers persists till 65 °C. Considering that for the other metal cations such band vanishes around 45 °C, we assume a higher stability for **1a**-$In^{3+}$ complexes and therefore presumably a higher binding constant.

Additionally, with the help of Van't Hoff plots (see ESI), it is possible to evaluate the thermodynamic parameters associated with the denaturation process. As a matter of fact, as the temperature increases the supramolecular structures based on PBI agglomerates coordinated to metal ions tend to disaggregate in solution. In details, the $F/F_0$ ratio can be considered proportional to the equilibrium constant of the process ($K_{eq}$) and a plot of $\ln(F/F_0)$ *vs.* $T^{-1}$ can thus provide information about the standard enthalpy ($\Delta H^o$) and entropy ($\Delta S^o$) of the denaturation process for each ion. As showed in the supporting information, the plots $\ln(F/F_0)$ *vs.* $T^{-1}$ are linear only at temperatures above 45°C. Therefore, we assume that both parameters are temperature dependent and decided to extrapolate their values only in the temperature interval 45°C ≤ T ≤ 75°C. With the exception of $Ce^{3+}$ and $Sm^{3+}$ $\Delta H^o$ can be determined to 54.1 ± 1.5 kJ·mol$^{-1}$ and $\Delta S^o$ 15.9 ± 2.7 J·mol$^{-1}$. For the former ions, $\Delta H^o$ can be calculated 16.2 ± 1.3 kJ·mol$^{-1}$ and 31.9 ± 2.1 kJ·mol$^{-1}$ and $\Delta S^o$ 2.2 ± 0.3 J·mol$^{-1}$ and 9.2 ± 0.4 J·mol$^{-1}$, respectively. In the case of $In^{3+}$ no calculation was carried out due to the fact that the denaturation process appears not to take place in the temperature range studied. The positive enthalpy and entropy values indicate that the denaturation of the metal-induced aggregates of **1a** in solution is an endothermic process and therefore favored at high temperature.

So far, we have described the complexation of **1a** in aqueous conditions, where **1a** is aggregated and the introduction of trivalent metal ions most likely trigger the formation of PBI emitting aggregates in solution.

**Fig. 7** Temperature dependent fluorescence spectra of **1a** in water after addition of 10 eq. of $M^{3+}$.

In order to get more insights into the interaction of the EDTA-PBI with metal cations in solution, it has been decided to add the metal ions to **1a** in a water/DMSO mixture. As mentioned above, **1a** is predominantly present as monomer in DMSO and these conditions should provide a better visualization of the true interaction of a single PBI molecule with the metal cation in solution. DMSO/water systems have been very well studied in the literature.[83-86] Due to the complex interactions of water and DMSO molecules, which lead for example to a pronounced melting point depression for a DMSO molar fraction ($x_{DMSO}$) of about 0.33[87] it would be difficult to attempt a detailed description of the effect of metal ions on the PET of **1a** over the entire molar fraction range. Therefore, we preferred to focus our attention to the region of $0.8 \leq x_{DMSO} \leq 1$ (so called, DMSO-rich region). Moreover, in order to assure a sufficient solubility of the metal salts and to avoid as much as possible the hydrophobic effect, it has been decided to work in mixtures where $x_{DMSO} = 0.9$.

Five salts of transition metals were chosen to be investigated. Namely, three divalent ($Ni^{2+}$, $Co^{2+}$ and $Cu^{2+}$) and two trivalent ($Al^{3+}$ and $Fe^{3+}$) metal ions have been tested. These five metal cations were selected among those who should form the most stable complexes with IDA and EDTA moieties. First of all, absorption measurements have been performed to investigate the influence of these metal cations on the spectroscopic properties of **1a**. As the EDTA-PBI is mostly present as a monomer in solution, absorption spectroscopy can provide precious information about the metal complexation (Figure 8).

**Fig. 8** Absorption spectra of **1a** in DMSO/water mixture (9:1) after addition of 10 eq. of $M^{x+}$.

The addition of bivalent and trivalent metal cations to the EDTA-PBI solution, results generally in a global decrease of the intensity of the absorption bands and in a lower value of the $I_{529/493}$ ratio. As described before, a decrease in the latter parameter accounts for the formation of aggregates in solution. In particular, these effects are strictly dependent on the metal ion and were recorded to be maximal for $Co^{2+}$ and $Al^{3+}$. A further proof of the interaction between **1a** and the metal cations in DMSO-rich solutions is given by fluorescence spectroscopy measurements. In DMSO-rich solution **1a** is still mainly present as monomer in solution.

**Fig. 9** Comparison of the $F/F_0$ ratios for complexes of 1a with $M^{2+}$ and $M^{3+}$ in a DMSO/water (9:1) mixture.

Upon addition of divalent metal cations, quenching of the PBI's fluorescence is observed, most likely due to the electron transfer processes from the metal center to the aromatic core. The affinity for trivalent metal cations remains high, as well. In DMSO-rich solutions, however, the addition of trivalent cations is now correlated with an enhanced fluorescence of **1a** (Figure 9). The presence of predominant monomeric species in solution allows a successful interaction of the EDTA-PBI with the metal cations which lead to a drastic quenching of the PET process ($Al^{3+}$ EF ≈ 1.18; $Fe^{3+}$ EF ≈ 1.83).

## Conclusion

In this contribution the first representatives of a new class of perylene bisimide–based surfactants is reported. In particular, the synthesis, optical characterization and metal complexation abilities of PBI derivative **1a** were investigated in detail. Two reasonable synthetic pathways, leading to target structure **1a**, have been elaborated and discussed. The full optical characterization (UV/Vis and emission spectroscopy) has been reported both under aqueous and in organic conditions. In the former solvent a detailed study of the aggregation behavior has been presented on the basis of absorption as well as fluorescence data and zeta potential measurements, both under basic as well acidic conditions.

Moreover, the influence of the intra-molecular PET process on the spectroscopic properties of dye **1a** in DMSO solution has been described. It has also been shown that upon addition of acid the PET process can be suppressed and the peculiarity of the phenomenon has been discussed with regard to feasible applications. For example, it could be imagined to fully exploit this pH-driven fluorescence sensitivity to develop EDTA-PBI water soluble sensors. Furthermore, a detailed investigation of the complexation of metal cations both in water and DMSO/water mixtures has been offered. This study revealed the importance of the aggregation state of the EDTA-PBI on the complexation properties of **1a**. In aqueous conditions, where bulky aggregates prevail, only big alkaline metal cations lead the suppression of the PET process. Bi- and trivalent cations were responsible for a pronounced fluorescence quenching (QF). Additionally, the complexation of trivalent metal ions is characterized by the formation of emitting aggregate species (excimers), which modify profoundly the emission spectrum of the EDTA-PBI itself.

The origin and stability of these species were discussed by means of temperature dependent fluorescence spectroscopy and $In^{3+}$ has been observed to form the strongest complexes, among the metal cations tested. A strong affinity of **1a** for di- and trivalent metal ions has been demonstrated in DMSO-rich solutions as well. The former metal cations are involved in energy/electron transfers also with predominantly monomeric PBIs. The addition of the latter results, instead, in a strong fluorescence enhancement (EF).

The intriguing structure of derivatives **1a – c** which combines namely an electron deficient central aromatic structure (PBI core), a polycarboxylic backbone and a chelating periphery, renders them suitable for many applications. Among the most appealing ideas, it would be interesting to exploit the potential of **1a** for the aqueous exfoliation of carbon allotropes or other 2D-layered inorganic materials, such as molybdenum disulphide ($MoS_2$) or tungsten disulphide ($WS_2$). In the latter case, in fact, it is known that specific ions (like $Ni^{2+}$ or $Al^{3+}$) are used to create inclusion during the exfoliation process and help to hinder the re-stacking of the dispersed material.[88]

Finally, the results of the complexation study, collected both in aqueous and organic conditions, underline the striking affinity of **1a** for heavy trivalent metal ions. The identification of such a successful interaction between transition metal and lanthanide cations with EDTA-PBIs opens up the way to challenging application in water decontamination, *e.g.* the detection of hard metals which is of extreme environmental importance.

## Experimental section

**Materials and methods:** Reagents and solvents were purchased from Acros Organics, Sigma Aldrich and used without further purification. Moisture sensitive reactions were performed under $N_2$ atmosphere. $CH_2Cl_2$ was freshly distilled from $CaH_2$, THF from Na/benzophenone and DMF dried over 4Å molecular sieves. Chromatographic purifications were performed with silica gel from Merck (Kieselgel 60, 40 – 60 µm, 230 – 400 mesh ASTM) in standard glass columns. TLC was performed on aluminium sheets coated with silica gel Merck (F254). $^1$H and $^{13}$C NMR spectra were recorded with a Brucker AV500 (500 MHz for $^1$H and 125 MHz for $^{13}$C) spectrometer, a Jeol JNM EX 400 (400 MHz for $^1$H and 100 MHz for $^{13}$C) and a Jeol Brucker Avance 300 (300 MHz for $^1$H and 75 MHz for $^{13}$C) spectrometer. Chemical shifts are reported in ppm at room temperature (RT) by using $CDCl_3$ as the solvent and internal standard, unless otherwise indicated. Abbreviations used for splitting patterns are s = singlet, d = doublet, t = triplet, m = multiplet, dd = double doublet. IR spectra were recorded with a FT-IR Nicolet 5700 and a Brucker Tensor 27 (ATR) and an ASI React IRTM 1000 spectrometer. For UV/Vis spectra a Perkin Elmer Lambda 1050 was used. Fluorescence was measured with a Horiba Scientific Fluorolog-3 spectrometer with CCD detector. MALDI-TOF mass spectrometry was carried out on a Shimadzu AXIMA Confidence, $N_2$ UV laser (337 nm), 50 Hz (reflectron). The matrix used were 2',4',6'-trihydroxyacetophenone monohydrate (THAP) 2,5-dihydoxybenzoic acid (DHB) 3,5-dimethoxy-4-hydroxycinnamic acid (SIN) 2-[(2*E*)-3-(4-*tert*-Butylphenyl)-2-methylprop-2-enylidene] malonitrile (DCTB). ESI mass spectrometry was performed with an Agilent Technologies 1100 Series LC/MSD Trap-SL spectrometer equipped with an ESI source, hexapole filter and ionic trap and a Brucker maXis 4G. Zeta-potential measurements were carried out on a Malvern Zetasizer Nano system with irradiation from a 633nm He-Ne laser. The solution of **1a** with the metal cations were prepared by mechanically stirring the solutions with a Heidolph Unimax 1010 platform shaker. For elemental analyses, a CE instrument EA 1110 CHNS was used.

In the following the syntheses of the different PBI derivatives is presented. The synthesis of the amine precursors **2**, **11** and **17** is presented in the electronic supporting information. Compound **4** was prepared according to a slightly modified procedure adapted from Xue *et al*.[89] For the putrescine based derivative **1a**, the direct synthesis will be defined as **route i**, while the convergent synthesis as **route ii**.

***N*-bis-(4-aminobutane)-3,4,9,10-PBI (4):**

[**route i**]: A mixture of PTCDA (10 g, $2.6\cdot10^{-2}$ mol) and 1,4 diaminobutane (4 eq.) in toluene (250 mL) was stirred at reflux (110 °C) for 4 hours. After cooling down to room temperature, the mixture was filtered under *vacuum* and washed with toluene. The crude solid was then re-suspended in KOH 5 M (200 mL) and stirred for 15 hours at ambient temperature. Subsequently, the suspension was filtered and **4** was collected as a red-brownish solid, which was dried in *vacuum* (16.4 g, yield = 89 %).

$^1$H NMR (500 MHz, DMSO-$d_6$ + TFA): $\delta$ = 1.65-1.76 (8H, 2m, 4$CH_2$), 2.89 (4H, m, 2$CH_2CH_2NH_3^+$), 4.07 (4H, t, 2$NCH_2CH_2$), 7.74 (6H, t, 2$CH_2NH_3^+$), 8.30 (4H, d, ArH), 8.55 (4H, d, 4ArH) ppm

ESI-MS: m/z 533.3 $(M+H)^+$, 267.2 $(M + 2H)^{2+}$

IR (KBr disc): $\nu$ = 3350, 3300 (primary amine, –$NH_2$ stretching); 2927, 2849 (stretching –$CH_2$); 1690, 1653 (stretching C=O bisimide) $cm^{-1}$

***N*-bis-(tert-butyl-(2,2’-aminobutylazanediyl)-diacetate)- 3,4,9,10 PBI (3):**

[**route i**]: A mixture of **4** (280 mg, $5.3\cdot10^{-4}$ mol), acetonitrile (15 mL), DIPEA (10 eq.) and *tert*-butyl bromoacetate (8 eq.) was stirred at 60 °C for 24 hours. Once cooled down to room temperature, it was *vacuum* filtered and the crude solid was washed with acetonitrile and water. Subsequently the solid residue was dissolved in chloroform (5 mL) and hexane was added (100 mL). The mixture was stirred for 10 minutes at room temperature and then let stand for one night. The precipitate was filtered and dried under *vacuum*. **3** is isolated as a brown solid (70 mg, yield = 13 %).

[**route ii**]: precursor amine **2** (404 mg, $1.3\cdot10^{-3}$ mol), PTCDA (250 mg, $6.4\cdot10^{-4}$ mol), imidazole (868 mg) and zinc acetate (35 mg) were mixed and heated up to 110°C for 4 h. Afterwards, dichloromethane was added to the solid residue and column chromatography in ($CH_2Cl_2$/EtOH 98:2) was performed to isolate a red solid, **3** (404 mg, yield = 64 %).

$^1$H NMR (300 MHz, $CDCl_3$, 25 °C): $\delta$ = 1.44 (s, 36H, 12 x $CH_3$), 1.63 (quintuplet, $J$ = 7.0 Hz, 4H, 2 x $CH_2$), 1.79 (quintuplet, $J$ = 7.5 Hz, 4H, 2 x $CH_2$), 2.77 (t, $J$ = 7.6 Hz, 4H, 2 x $CH_2$), 3.44 (s, 8H, 4 x $NCH_2$), 4.21 (t, $J$ = 7.4 Hz, 4H, 2 x $CH_2$), 8.33 (d, $J$ = 8.0 Hz, 4H, ArH), 8.48 (d, $J$ = 8.0 Hz, 4H, ArH) ppm

$^{13}C$ NMR (75 MHz, $CDCl_3$, 25 °C): $\delta$ = 25.536 (2 C, $CH_2$), 25.641 (2 C, $CH_2$), 28.049 (12 C, $CH_3$), 40.267 (2 C, $CH_2$), 53.864 (2 C, $CH_2$), 55.732 (4 C, $CH_2$), 80.732 (4 C, quat. C $^t$Bu), 122.469 (4 C, Ar-CH), 122.850 (2 C, Ar-C), 125.388 (2 C, Ar-C), 128.552 (4 C, Ar-C), 130.625 (4 C, Ar-CH), 133.581 (4 C, Ar-C), 162.762 (4 C, CON), 170.773 (4 C, COO) ppm

MALDI-TOF (THAP): m/z 989 $(M+H)^+$, 1011 $(M + Na)^+$

IR (ATR): $\nu$ = 2976.44, 2932.02, 1731.53, 1692.97, 1654.17, 1594.27, 1340.47, 1251.21, 1215.23, 1142.65, 988.15, 809.48, 745.88 $cm^{-1}$

EA for $C_{56}H_{68}N_4O_{12}$: calcd. C 68.00, H 6.93, N 5.66; found C 67.66; H 6.90; N 5.65

***N*-bis-(tert-butyl-(2,2'-aminobutylazanediyl)-diacetic acid)- 3,4,9,10-PBI (1a):**

[**route i**]: A solution of **3** (520 mg, $5.3\cdot10^{-4}$ mol) in formic acid (20 mL) was stirred at room temperature for 2 days. Acetonitrile (20 mL) was added and a solid precipitated. The solvent was evaporated *in vacuo*. The crude solid was washed two times with acetonitrile and once with diethyl ether. **1a** is isolated as reddish solid (400 mg, quantitative yield).

[**route ii**]: A solution of **3** (250 mg, $2.5\cdot10^{-4}$ mol) in TFA:$CH_2Cl_2$ (1:1) was stirred at room temperature for 5 days. After evaporation of the solvent, the product was precipitated by addition of diethylether. The solid was filtrated and dried *in vacuo*. A red-brown solid was obtained (160 mg, yield = 84 %).

$^1H$ NMR (300 MHz, $CDCl_3$, 25 °C): $\delta$ = 1.99-2.06 (m, 8H, 4 x $CH_2$), 3.69 (t, $J$ = 7.4 Hz, 4H, 2 x $CH_2$), 4.39 (t, $J$ = 6.6 Hz, 4H, 2 x $CH_2$), 4.45 (s, 8H, 4 x $NCH_2$), 8.80-8.86 (m, 8H, perylene ArH) ppm

$^{13}C$ NMR (75 MHz, $CDCl_3$, 25 °C): $\delta$ = 22.080 (2 C, $CH_2$), 24.631 (2 C, $CH_2$), 40.479 (2 C, $CH_2$), 55.547 (2 C, $CH_2$), 58.388 (4 C, $CH_2$), 122.430 (2 C, Ar-C), 124.888 (4 C, Ar-CH), 126.837 (2 C, Ar-C), 129.781 (4 C, Ar-C), 133.566 (4 C, Ar-CH), 136.483 (4 C, Ar-C), 166.146 (4 C, CON), 169.575 (4 C, COO) ppm

MALDI-TOF (DHB): m/z 649 $(MH - 2CH_2CO_2)^+$, 708 $(MH - CH_2CO_2)^+$, 765 $(M + H)^+$, 787 $(M + Na)^+$

IR (ATR): $\nu$ = 3468.87, 3016.23, 2969.14, 2547.61, 1735.46, 1687.25, 1645.08, 1593.03, 1576.71, 1441.85, 1402.08, 1381.67, 1341.39, 1246.11, 1169.21, 1137.12, 1088.06, 809.08, 794.31, 744.42, 719.63 $cm^{-1}$

EA for $C_{46}H_{39}F_9N_4O_{18}$ (765) x 3 $CF_3COOH$: calcd. C 49.92, H 3.55, N 5.06; found C 49.37; H 4.10; N 4.96

**tetra-tert-butyl 2,2',2'',2'''-(((((1,3,8,10-tetraoxoanthra[2,1,9-def:6,5,10-d'e'f']di-isoquinoline-2,9(1H,3H,8H,10H)-diyl)bis(propane-3,1diyl)-bis((2-(tert-butoxy)-2-oxoethyl)azanediyl))bis(butane-4,1-diyl))bis(azanetriyl)-tetraacetate (5):**

Precursor amine **11** (0.62 g, 1.3 mmol), PTCDA (0.25 g, 0.64 mmol), imidazole (0.87 g, 13.0 mmol) and zinc acetate (0.035 g, 0.2 mmol) were heated at 110 °C for 4 h. Afterwards, dichloromethane was added and the solid residue was purified by column chromatography ($SiO_2$, dichloromethane:ethanol, 95:5). **5** is isolated as red solid (gummy), (1.9 g, yield = 53.3 %).

$^1$H NMR (300 MHz, $CDCl_3$): $\delta$ = 1.43 (s, 18H, 6 x $CH_3$), 1.44 (bs, 36H, 12 x $CH_3$) 1.47-1.48 (m, 4H, 2 x $CH_2$), 1.74 (broad quintuplet, 4H, 2 x $CH_2$), 1.92 (quintuplet, *J* = 7.2 Hz, 4H, 2 x $CH_2$), 2.63 (t, *J* = 6.8 Hz, 4H, 2 x $CH_2$), 2.68 (t, *J* = 7.0 Hz, 4H, 2 x $CH_2$), 2.78 (t, *J* = 7.0 Hz, 4H, 2 x $CH_2$), 3.30 (s, 4H, 2 x $NCH_2$), 3.41 (s, 8H, 4 x $NCH_2$), 4.24 (t, *J* = 7.6 Hz, 4H, 2 x $CH_2$), 8.51 (d, *J* = 8.0 Hz, 4H, ArH), 8.61 (d, *J* = 8.0 Hz, 4H, ArH) ppm

$^{13}$C NMR (75 MHz, $CDCl_3$): $\delta$ = 25.18 (2 C, $CH_2$), 25.66 (2 C, $CH_2$), 25.87 (2 C, $CH_2$), 28.04 (6 C, $CH_3$), 28.07 (12 C, $CH_3$), 38.91 (2 C, $CH_2$), 51.76 (2 C, $CH_2$), 53.73 (2 C, $CH_2$), 54.06 (2 C, $CH_2$), 55.14 (2 C, $CH_2$), 55.72 (4 C, $CH_2$), 80.56 (2 C, quat. C $^t$Bu), 80.70 (4 C, quat. C $^t$Bu), 122.71 (4 C, Ar-CH), 123.07 (2 C, Ar-C), 125.83 (2 C, Ar-C), 128.90 (4 C, Ar-C), 130.91 (4 C, Ar-CH), 133.99 (4 C, Ar-C), 162.99 (4 C, CON), 170.79 (4 C, COO), 170.95 (2 C, COO) ppm

MS-ESI(+): *m/z* = 1332 [$M^+$ + H]

IR (ATR): *v* = 2975.77, 2933.03, 2865.49, 1729.00, 1694.37, 1655.12, 1594.05, 1440.45, 1402.63, 1365.62, 1345.19, 1247.93, 1216.78, 1145.75, 1069.42, 848.50, 809.39, 744.38 cm$^{-1}$

EA for $C_{74}H_{102}N_6O_{16}$: calcd. C 66.74, H 7.72, N 6.31; found C 66.30; H 7.80; N 6.35

**2,2',2'',2'''-(((((1,3,8,10-tetraoxoanthra[2,1,9-def:6,5,10-d'e'f']diisoquinoline-2,9(1H,3H,8H,10H)-diyl)bis(propane-3,1diylbis((carboxymethyl)azanediyl-bis(butane-4,1-diyl))bis(azanetriyl))tetraacetic acid (1b):**

**5** (0.5 g, 0.37 mmol) was dissolved in 18 mL of TFA. The reaction mixture was stirred for 3 days at room temperature. After evaporation of the solvent, the product was precipitated on addition of diethyl ether. After filtration, the product was dried under *vacuum*. **1b** is isolated as dark red solid (0.30 g, yield = 81.9%).

$^1$H NMR (300 MHz, TFA:$CDCl_3$; (1:1)): $\delta$ = 1.85 (bq, 8H, 4 x $CH_2$), 2.25 (bq, *J* = 6.8 Hz, 4H, 2 x $CH_2$), 3.33-3.46 (m, 12H, 6 x $CH_2$), 4.02-4.30 (m, 12H of 6 x $NCH_2$ superimposed with 4H protons of 2 x $CH_2$), 8.61 (d, *J* = 7.6 Hz, 4H, perylene ArH), 8.67 (d, *J* = 8.4 Hz, 4H, perylene ArH) ppm

MS-ESI(+): *m/z* = 996 [$M^+$ + 2H]

IR (ATR): *v* = 3016.80, 2973.38, 2546.03, 1735.01, 1691.09, 1649.09, 1593.05, 1577.36, 1401.75, 1343.23, 1172.33, 1129.89, 809.29, 794.60, 719.90 cm$^{-1}$

EA for $C_{50}H_{54}N_6O_{16}$ x 5 $CF_3COOH$: calcd. C 46.04, H 3.80, N 5.37; found C 45.74; H 4.29; N 5.07

**tetra-tert-butyl 2,2',2'',2'''-(((11,11'-((1,3,8,10-tetraoxoanthra[2,1,9-def:6,5,10-d'e'f']diisoquinoline-2,9(1H,3H,8H,10H)-diyl)bis(propane-3,1-diyl-bis(2,2,15,15-tetramethyl-4,13-dioxo-3,14-dioxa-6,11-diazahexadecane-11,6-diyl-bis(propane-3,1-diyl))bis(azanetriyl))tetraacetate (6):**

Precursor amine **17** (0.25 g, 0.38 mmol), PTCDA (0.07 g, 0.17 mmol), imidazole (0.24 g, 3.46 mmol) and zinc acetate (0.01 g, 0.05 mmol) were heated at 110°C for 4 h. Afterwards, dichloromethane was added to the solid residue and residue was purified through column chromatography ($SiO_2$, dichloromethane:ethanol, 95:5). **6** is isolated as dark red solid (0.22 g, yield = 74.3 %).

$^1$H NMR (300 MHz, $CDCl_3$): $\delta$ = 1.41-1.42 (s, 72H, 24 x $CH_3$. This multiplet superimpose on the signals of 8H protons of 4 x $CH_2$), 1.61 (quintuplet, $J$ = 6.9 Hz, 4H, 2 x $CH_2$), 1.91 (quintuplet, $J$ = 7.0 Hz, 4H, 2 x $CH_2$), 2.57-2.60 (m, 12H, 6 x $CH_2$), 2.68 (t, $J$ = 7.4 Hz, 4H, 2 x $CH_2$), 2.77 (t, $J$ = 7.0 Hz, 4H, 2 x $CH_2$), 3.20 (s, 4H, 2 x $NCH_2$), 3.28 (s, 4H, 2 x $NCH_2$), 3.40 (s, 8H, 4 x $NCH_2$), 4.22 (t, $J$ = 7.4 Hz, 4H, 2 x $CH_2$), 8.43 (d, $J$ = 8.0 Hz, 4H, ArH), 8.55 (d, $J$ = 8.0 Hz, 4H, ArH) ppm

$^{13}$C NMR (75 MHz, $CDCl_3$): $\delta$ = 25.219 (2 C, $CH_2$), 25.372 (2C, $CH_2$), 25.884 (2 C, $CH_2$), 26.056 (2 C, $CH_2$), 28.023 (12 C, $CH_3$), 28.040 (6 C, $CH_3$), 28.064 (6 C, $CH_3$), 38.906 (2 C, $CH_2$), 51.789 (2 C, $CH_2$), 51.885 (2 C, $CH_2$), 52.113 (2 C, $CH_2$), 53.805 (2 C, $CH_2$), 54.220 (2 C, $CH_2$), 55.191 (2 C, $CH_2$), 55.519 (2 C, $CH_2$), 55.704 (4 C, $CH_2$), 80.482 (4 C, quat. C $^t$Bu), 80.523 (2 C, quat. C $^t$Bu), 80.684 (2 C, quat. C $^t$Bu), 122.776 (4 C, Ar-CH), 123.101 (2 C, Ar-C), 125.908 (2 C, Ar-C), 128.952 (4 C, Ar-C), 130.956 (4 C, Ar-CH), 134.072 (4 C, Ar-C), 163.032 (4 C, CON), 170.728 (4 C, COO), 170.923 (4 C, COO) ppm

MS-ESI(+): $m/z$ = 1675 [$M^+$ + 2H]

IR (ATR): $v$ = 2976.98, 2935.25, 1727.43, 1695.03, 1655.53, 1594.39, 1365.73, 1247.92, 1216.67, 1144.99, 848.11, 809.43, 744.48 $cm^{-1}$

EA for $C_{92}H_{136}N_8O_{20}$: calcd. C 66.00, H 8.19, N 6.69; found C 64.87; H 8.65; N 6.56

**2,2',2'',2'''-((((((1,3,8,10-tetraoxoanthra[2,1,9-def:6,5,10-d'e'f']diisoquinoline-2,9(1H,3H,8H,10H)-diyl)bis(propane-3,1-diyl-bis((carboxymethyl)azanediyl-bis(butane-4,1-diyl-bis((carboxymethyl)azanediyl))bis(propane-3,1-diyl-**

**bis(azanetriyl))tetraacetic acid (1c):**

**6** (0.23 g, 0.14 mmol) was dissolved in 18 mL of TFA. The reaction mixture was stirred for 6 days at RT. After evaporation of the solvent, the product was precipitated on addition of diethyl ether. After filtration, the product was dried under *vacuum*. **1c** is isolated as dark red solid (9.4 mg, yield = 56.2%).

$^1H$ NMR (300 MHz, TFA:$CDCl_3$; (1:1)): $\delta$ = 1.99 (broad quintuplet, 8H, 4 x $CH_2$), 2.41 (broad quintuplet, 4H, 2 x $CH_2$), 2.50 (broad quintuplet, 4H, 2 x $CH_2$), 3.49 (bt, 8H, 4 x $CH_2$), 3.57 (bt, 8H, 4 x $CH_2$), 3.68 (bt, 8H, 4 x $CH_2$), 4.21 (s, 4H, 2 x $NCH_2$), 4.39 (s, 12H, 6 x $NCH_2$), 8.78 (d, $J$ = 6.0 Hz, 4H, ArH), 8.84 (d, $J$ = 6.0 Hz, 4H, ArH) ppm

MS-ESI(+): $m/z$ = 1226 [$M^+$ + 2H]

IR (ATR): $\nu$ = 2976.98, 2935.25, 1727.43, 1695.03, 1655.53, 1594.39, 1365.73, 1247.92, 1216.67, 1144.99, 848.11, 809.43, 744.48 $cm^{-1}$

EA for $C_{60}H_{72}N_8O_{20}$ x 5 $CF_3COOH$: calcd. C 46.83, H 4.32, N 6.24; found C 45.66; H 4.38; N 6.12

## Acknowledgements

MM, FH and AH thank the Deutsche Forschungsgemeinschaft (DFG - SFB 953, Project A1 “Synthetic Carbon Allotropes”) and the Interdisciplinary Center for Molecular Materials (ICMM) for the financial support. The research leading to these results has partially received funding from the European Union Seventh Framework Programme under grant agreement n°604391 Graphene Flagship. PS is thankful to Alexander von Humboldt foundation for granting the AvH research Fellowship.

**Supporting Information**

# Novel EDTA-Ligands Containing an Integral Perylene Bisimide (PBI) Core as Optical Reporter Units

Mario Marcia, Prabhpreet Singh, Frank Hauke, Michele Maggini and Andreas Hirsch*

1. Synthesis of precursor amine **2**
2. Synthesis of precursor amine **11**
3. Synthesis of precursor amine **17**
4. Solubility tests for **1a**
5. NMR Study for the water induced aggregation of **1a** in DMSO solutions
6. Fluorescence data for the complexation of metal ions by **1a** in water solutions

## Synthesis of di-tert-butyl 2,2'-((4-(1,3-dioxoisoindolin-2-yl)butyl)azanediyl)diacetate (2)

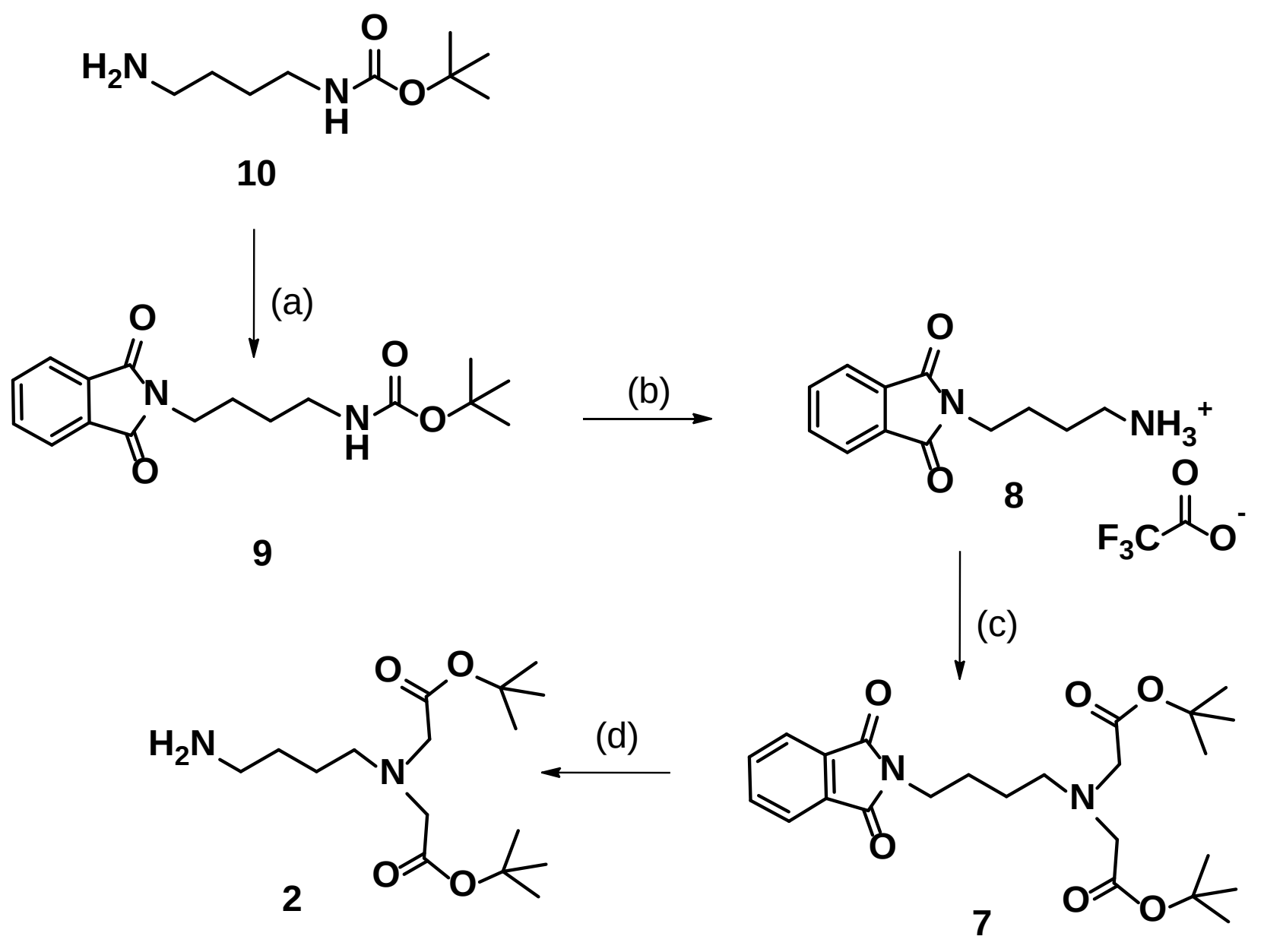

**Scheme S1.** Synthesis of precursor amine **2**. (a) Phthalic anhydride, 135°C, 10 min; (b) trifluoroacetic acid, $CH_2Cl_2$, 4 h, RT; (c) *tert*-butyl 2-bromoacetate, 1,4-dioxan, DIEA, 60°C, 24h; (d) hydrazine hydrate, ethanol, 50°C, 6 h.

Compound **10** was synthesized according to Muller *et al.* [1]

*Synthesis of tert-Butyl 4-(1,3-dioxoisoindolin-2-yl)butylcarbamate (**9**)*:
1.00 g of **10** (5.30 mmol) was reacted with phthalic anhydride (0.80 g, 5.30 mmol) at 135°C for 10 min under solvent free conditions. After cooling the reaction mixture, methanol was added to remove potentially formed side products by filtration. The filtrate was concentrated under vacuum and **9** was purified by column chromatography ($SiO_2$, dichloromethane/methanol), yielding a white solid; yield 1.22 g (3.85 mmol, 72.6 %); $R_f$ = 0.7 (dichloromethane/methanol 98:2).
$^1$H NMR (300 MHz, $CDCl_3$, 25°C): δ = 1.41 (s, 9H, 3 x $CH_3$), 1.51 (quintuplet, $J$ = 7.6 Hz, 2H, $CH_2$), 1.69 (quintuplet, $J$ = 7.2 Hz, 2H, $CH_2$), 3.14 (quartet, $J$ = 6.4 Hz, 2H, $CH_2$), 3.69 (t, $J$ = 7.2 Hz, 2H, $CH_2$), 4.58 (bs, 1H, NH of carbmate), 7.70 (dd, $J_1$ = 3.2 Hz, $J_2$ = 5.2 Hz, 2H, ArH), 7.83 (dd, $J_1$ = 2.8 Hz, $J_2$ = 5.2 Hz, 2H, ArH) ppm.
$^{13}$C NMR (75 MHz, $CDCl_3$, 25 °C): δ = 25.78 (1 C, $CH_2$), 27.25 (1 C, $CH_2$), 28.21 (3 C, $CH_3$), 37.39 (1 C, $CH_2$), 39.88 (1 C, $CH_2$), 78.94 (1 C, quat. C $^t$Bu), 123.13 (2 C, Ar-CH), 132.02 (2 C, Ar-C), 133.88 (2 C, Ar-CH), 155.92 (2 C, CON), 168.37 (1 C, CO(NH)O) ppm.
MS-ESI(+): $m/z$ = 318 [$M^+$], 217 [$M^+$ - Boc].
IR (ATR): ν = 3371.86, 2978.38, 2934.74, 1701.97, 1679.91, 1527.06, 1398.11, 1361.34, 1310.39, 1271.40, 1247.20, 1170.13, 1051.06, 1012.83, 717.48 $cm^{-1}$.

*Synthesis of 2-(4-aminobutyl)isoindoline-1,3-dione (**8**)*:
To a solution of **9** (3.00 g, 9.40 mmol) in 5 mL dicholoromethane 5 mL of trifluoroacetic acid (TFA) was added to cleave the Boc protection group. The reaction mixture was stirred for 4 h at RT. After evaporation of the solvent, the product was precipitated with diethyl ether. After filtration, the product was dried under vacuum yielding compound **8** as TFA salt, white solid, quantitative yield; $R_f$ = 0.35 (dichloromethane/methanol 90:10).

$^1$H NMR (300 MHz, $CDCl_3$, 25 °C): 1.53 (quintuplet, *J* = 8.8 Hz, 2H, $CH_2$), 1.62 (quintuplet, *J* = 6.4 Hz, 2H, $CH_2$), 2.80 (sextet, *J* = 6.4 Hz, 2H, $CH_2NH_3^+$, converts to triplet on $D_2O$ exchange), 3.60 (t, *J* = 6.4 Hz, 2H, $CH_2$), 7.73 (bs, 3H, $NH_3^+$ exchanges with $D_2O$), 7.83-7.89 (m, 4H, ArH) ppm.
$^{13}$C NMR (75 MHz, $CDCl_3$, 25 °C): δ = 24.39 (1 C, $CH_2$), 24.94 (1 C, $CH_2$), 36.75 (1 C, $CH_2$), 38.34 (1 C, $CH_2$), 123.05 (2 C, Ar-CH), 131.61 (2 C, Ar-C), 134.47 (2 C, Ar-CH), 168.06 (2 C, CON) ppm.
MS-ESI(+): *m/z* = 219 [$M^+$] as free amine.
IR (ATR): ν = 3400.15, 2937.11, 1707.96, 1676.22, 1398.49, 1200.59, 1179.01, 1127.86, 1065.16, 800.49, 721.48, 711.19 $cm^{-1}$.

*Synthesis of tert-Butyl 2,2'-(4-(1,3-dioxoisoindolin-2-yl)butylazanediyl)diacetate (**7**)*:
To a solution of **8** (1.40 g, 4.20 mmol) in 1,4-dioxane (20 mL), *tert*-butyl 2-bromoacetate (1.43 mL, 8.80 mmol) and diisopropylethyl amine (DIEA) (2.23 mL, 12.65 mmol) was added. The reaction mixture was stirred for 24 h at 60°C. After completion of the reaction (tlc), the reaction mixture was filtered and the solvent was evaporated under vacuum. Purificatio by column chromatography ($SiO_2$, dichloromethane/methanol) was applied to isolate pure **7** as a liquid, yield 1.54 g (3.45 mmol, 81.9 %); $R_f$ = 0.55 (dichloromethane/methanol 99:1).
$^1$H NMR (300 MHz, $CDCl_3$, 25 °C): δ = 1.43 (s, 18H, 6 x $CH_3$), 1.47-1.54 (m, 2H, $CH_2$), 1.70 (quintuplet, *J* = 7.4 Hz, 2H, $CH_2$), 2.71 (t, *J* = 7.4 Hz, 2H, $CH_2$), 3.40 (s, 4H, 2 x $NCH_2$), 3.69 (t, *J* = 7.0 Hz, 2H, $CH_2$), 7.70 (dd, $J_1$ = 2.8 Hz, $J_2$ = 5.2 Hz, 2H, ArH), 7.82 (dd, $J_1$ = 2.8 Hz, $J_2$ = 5.2 Hz, 2H, ArH) ppm.
$^{13}$C NMR (75 MHz, $CDCl_3$, 25 °C): δ = 25.18 (1 C, $CH_2$), 26.07 (1 C, $CH_2$), 27.99 (6 C, $CH_3$), 37.66 (1 C, $CH_2$), 53.49 (1 C, $CH_2$), 55.75 (2 C, $CH_2$), 80.76 (2 C, quat. C $^t$Bu), 123.09 (2 C, Ar-CH), 132.13 (2 C, Ar-C), 133.81 (2 C, Ar-CH), 168.395 (2 C, CON), 170.683 (2 C, COO) ppm.
MS-ESI(+): *m/z* = 446 [$M^+$].
IR (ATR): ν = 2976.93, 2934.60, 1771.72, 1708.64, 1394.13, 1366.48, 1218.25, 1145.63, 751.54 $cm^{-1}$.

*Synthesis of tert-Butyl 2,2'-(4-aminobutylazanediyl)diacetate (**2**)*:
To a solution of **7** (3.40 g, 2.00 mmol) in ethanol (35 mL), hydrazine hydrate (0.92 mL, 4.00 mmol) was added. The reaction mixture was stirred for 6 h at 50°C. The formed precipitate was removed by filtration. The filtrate was concentrated in *vacuo* and the residue was dissolved in dichloromethane and washed with 10% KOH solution. The organic layer was back extracted with brine (50 mL) and dried over $Na_2SO_4$ and concentrated in vacuum and purified by column chromatography ($SiO_2$, dichloromethane/methanol/triethylamine) to isolate pure **2** as a liquid, yield 1.63 g (5.16 mmol, 67.8 %); $R_f$ = 0.65 (dichloromethane:methanol:triethylamine 80:15:5).
$^1$H NMR (300 MHz, $CDCl_3$, 25 °C): δ = 1.44-1.45 (m, 18H, 6 x $CH_3$); the broad signals of the Boc group protons superimpose the signals of the putrescine methylene protons (4H, 2 x $CH_2$), 2.66-2.70 (m, 4H, 2 x $CH_2$), 3.41 (s, 4H, 2 x $NCH_2$) ppm.
$^{13}$C NMR (75 MHz, $CDCl_3$, 25 °C): δ = 25.12 (1 C, $CH_2$), 27.96 (6 C, $CH_3$), 30.94 (1 C, $CH_2$), 41.73 (1 C, $CH_2$), 53.77 (1 C, $CH_2$), 55.77 (2 C, $CH_2$), 80.76 (2 C, quat. C $^t$Bu), 170.70 (2 C, COO) ppm.
MS-ESI(+): *m/z* = 316 [$M^+$].

## Synthesis of tert-butyl 2,2'-(4-((3-aminopropyl)(2-tert-butoxy-2-oxoethyl)amino)-butylazanediyl) diacetate (11)

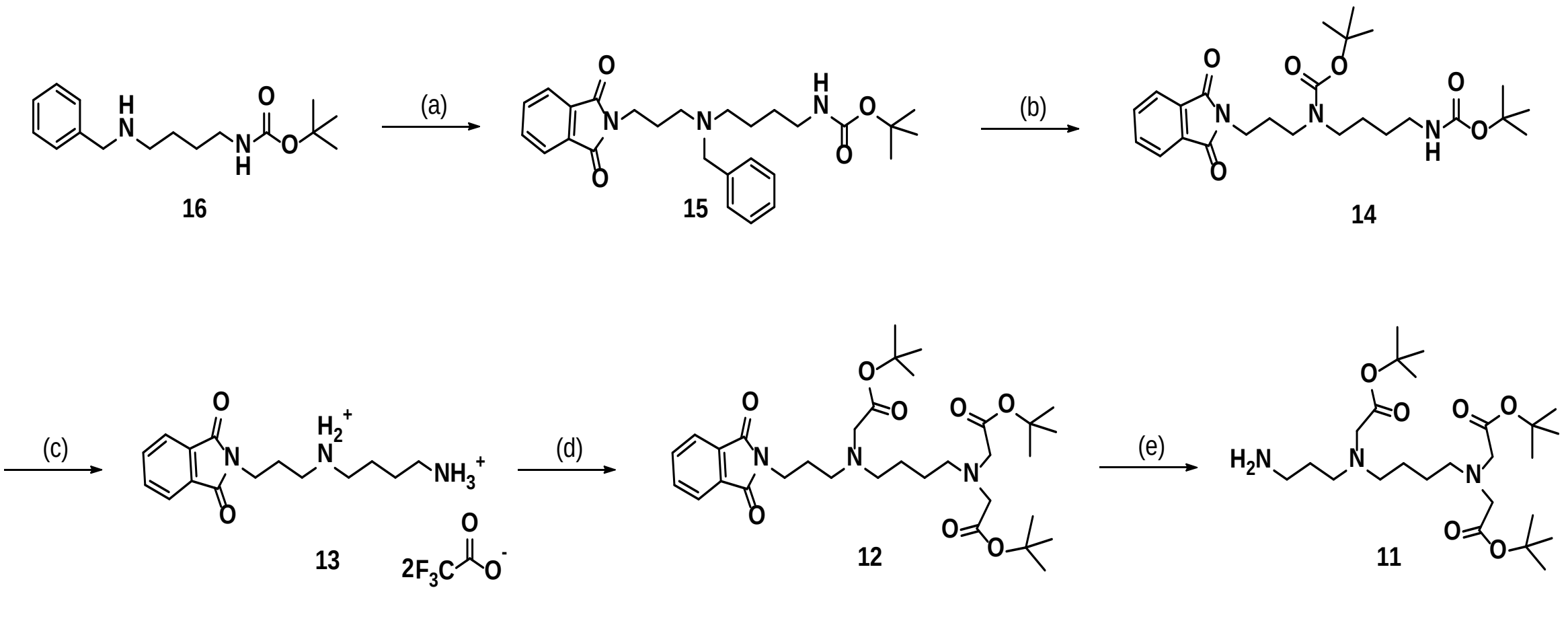

**Scheme S2.** Synthesis of precursor amine **11**. (a) *N*-(3-bromopropyl)phthalimide, chloroform:acetonitrile (1:1), $Na_2CO_3$, 80°C, 3 days; (b) 10% Pd/C, $H_2$, di-tert-butyl dicarbonate, acetic acid (cat.), $CH_3OH$; (c) trifluoroacetic acid, $CH_2Cl_2$, 4 h, RT; (d) *tert*-butyl 2-bromoacetate, 1,4-dioxan, DIEA, 60°C, 24 h; (e) hydrazine hydrate, ethanol, 50°C, 6 h.

*Synthesis of tert-butyl 4-(benzylamino)butylcarbamate (**16**)*:
**10** (6.00 g, 31.00 mmol) was dissolved in 60 mL of methanol. To this solution benzaldehyde (4.06 g, 38.00 mmol), $MgSO_4$ (7.70 g, 63.00 mmol) and triethylamine (0.97 g, 9.00 mmol) was added and the mixture was stirred at RT overnight. Afterwards, the reaction mixture was cooled to 0°C and $NaBH_4$ (6.00 g, 160.00 mmol) was added over a period of 60 minutes. The reaction mixture was stirred for additional 60 minute at 0°C and then at room temperature for 2 hours. Water (200 mL) was added to quench the reaction and the aqueous layer was extracted with ethyl acetate (3 x 150 mL). The organic layer was dried over $Na_2SO_4$ and afterwards, the solvent was evaporated under vacuum. The residue was purified by column chromatography ($SiO_2$, dichloromethane:methanol) yielding **16** as a liquid, yield 6.86 g (21.5 mmol, 67.6 %); $R_f$ = 0.5 dichloromethane:methanol (90:10).
$^1$H NMR (300 MHz, $CDCl_3$, 25 °C): $\delta$ = 1.41 (s, 9H, 3 x $CH_3$), 1.51 (quintuplet, $J$ = 3.2 Hz, 4H, 2 x $CH_2$), 1.69 (bs, 1H, NH exchanges with $D_2O$), 2.62 (bt, $J$ = 6.8 Hz, 2H, $CH_2$), 3.09 (bq, $J$ = 5.2 Hz, 2H, $CH_2$, changes to broad triplet on $D_2O$ exchange), 3.75 (s, 2H, $NCH_2$), 4.85 (bs, 1H, NH of carbmate), 7.21-7.30 (m, 5H, ArH) ppm.
$^{13}$C NMR (75 MHz, $CDCl_3$, 25 °C): $\delta$ = 27.018 (1 C, $CH_2$), 27.584 (1 C, $CH_2$), 28.218 (3 C, $CH_3$), 40.073 (1 C, $CH_2$), 48.580 (1 C, $CH_2$), 53.649 (1 C, $CH_2$), 78.750 (1 C, quat. C $^t$Bu), 126.870 (1 C, Ar-CH), 128.063 (2 C, Ar-CH), 128.30 (2 C, Ar-CH), 140.023 (1 C, Ar-C), 155.990 (1 C, CO(NH)O).
MS-ESI(+): *m*/*z* = 279 [M$^+$+H].
IR (ATR): $\nu$ = 3338.51, 2975.05, 2930.46, 2862.96, 1690.45, 1514.38, 1452.98, 1390.67, 1364.51, 1270.63, 1248.79, 1167.66, 733.61, 697.56 cm$^{-1}$.

*Synthesis of tert-butyl 4-(benzyl(3-(1,3-dioxoisoindolin-2-yl)propyl)amino)butylcarbamate (**15**)*:
To a solution of **16** (5.0 g, 18.00 mmol) in acetonitrile:chloroform ($CH_3CN$:$CHCl_3$) (1:1) (100 mL), *N*-(3-bromopropyl)phthalimide (6.26 g, 23.00 mmol) and $Na_2CO_3$ (3.8 g, 36.00 mmol) was added. The reaction mixture was stirred for 3 days at 80°C. After completion of the reaction (tlc), the reaction mixture was filtered and the solvent was evaporated under vacuum. The residue was purified by column chromatography ($SiO_2$, dichloromethane:methanol) yielding **15** as transparent light greenish liquid (slowly become solid when kept in refrigerator), yield 6.83 g (14.7 mmol, 81.7 %); $R_f$ = 0.75 dichloromethane:methanol (97:3).

$^1$H NMR (300 MHz, $CDCl_3$, 25 °C): δ = 1.42 (s, 9H, 3 x $CH_3$); 1.47 (quintuplet, $J$ = 3.2 Hz, 4H, 2 x $CH_2$), 1.82 (quintuplet, $J$ = 7.2 Hz, 2H, $CH_2$), 2.41 (bt, $J$ = 6.4 Hz, 2H, $CH_2$), 2.46 (t, $J$ = 6.8 Hz, 2H, $CH_2$), 3.07 (bq, $J$ = 5.6 Hz, 2H, $CH_2$), 3.52 (s, 2H, $NCH_2$), 3.68 (t, $J$ = 7.6 Hz, 2H, $CH_2$), 4.78 (bs, 1H, NHCO), 7.16-7.30 (m, 5H, ArH-benzyl), 7.70 (dd, $J_1$ = 3.2 Hz, $J_2$ = 5.6 Hz, 2H, ArH), 7.82 (dd, $J_1$ = 3.2 Hz, $J_2$ = 5.6 Hz, 2H, ArH) ppm.
$^{13}$C NMR (75 MHz, $CDCl_3$, 25 °C): δ = 24.285 (1C, $CH_2$), 25.898 (1C, $CH_2$), 27.629 (1 C, $CH_2$), 28.278 (3C, $CH_3$), 36.251 (1 C, $CH_2$), 40.327 (1 C, $CH_2$), 50.988 (1 C, $CH_2$), 53.203 (1 C, $CH_2$), 58.454 (1 C, $CH_2$) 78.750 (1 C, quat. C $^t$Bu), 123.087 (2 C, Ar-CH), 126.755 (1 C, Ar-CH), 128.089 (1 C, Ar-CH), 128.834 (1 C, Ar-CH), 132.123 (2 C, Ar-C), 133.816 (2 C, Ar-CH), 139.514 (1 C, Ar-C), 156.027 (2 C, CON), 168.384 (1C, CO(NH)O).
MS-ESI(+): $m/z$ = 466 [$M^+$+H] IR (ATR).
ν = 3389.68, 2944.63, 2867.67, 2799.57, 1703.53, 1685.33, 1522.87, 1390.90, 1365.22, 1270.42, 1243.89, 1172.29, 741.30, 721.08 $cm^{-1}$.

*Synthesis of tert-butyl 4-((3-(1,3-dioxoisoindolin-2-yl)propyl)(tert-butoxycarbonyl)amino)butyl-carbamate (**14**)*:
To the $N_2$-flushed solution of **15** (3.00 g, 6.40 mmol) in 30 mL of methanol (dry), 10% Pd/C (0.45 g) was added, followed by the addition of acetic acid (0.25 mL). Di-*tert*-butyldicarbonate (2.26 g, 10.00 mmol) was also added to the above solution. The flask was degassed and saturated with hydrogen and stirred for 24 h at room temperature. The reaction progress was monitored by TLC. After completion of the reaction (tlc), the Pd/C was filtered off (celite), washed with methanol and the combined organic solvents were evaporated under vacuum. The residue was purified by column chromatography ($SiO_2$, dichloromethane methanol) yielding **14** as liquid, yield 2.60 g (5.47 mmol, 84.9 %); $R_f$ = 0.65 in 2% methanol:dichloromethane.
$^1$H NMR (300 MHz, $CDCl_3$, 25 °C): δ = 1.41 [(bs, 18H, $CH_3$) which splits into two singlets at 1.35 (s, 9H, 3 x $CH_3$) and 1.38 (s, 9H, 3 x $CH_3$) when the NMR was taken at -20$^o$C], 1.45-1.47 (m, 2H, $CH_2$), 1.53 (quintuplet, $J$ = 7.4 Hz, 2H, $CH_2$), 1.89 (quintuplet, $J_1$ = 7.2 Hz, 2H, $CH_2$), 3.11 (q, $J$ = 6.0 Hz, 2H, $CH_2$), 3.20-3.26 [(broad, splits into two triplets at 3.16 (t, $J$ = 7.2 Hz, 2H, $CH_2$) and 3.25 (t, $J$ = 7.2 Hz, 2H, $CH_2$) when the NMR was accquired at -20$^o$C], 3.68 (t, $J$ = 7.2 Hz, 2H, $CH_2$), 4.67 (bs, 1H, NH of carbmate), 7.71 (dd, $J_1$ = 2.8 Hz, $J_2$ = 5.2 Hz, 2H, ArH), 7.83 (dd, $J_1$ = 2.8 Hz, $J_2$ = 5.2 Hz, 2H, ArH) ppm.
$^{13}$C NMR (75 MHz, $CDCl_3$, 25 °C): δ = 25.502 (1 C, $CH_2$), 27.214 (1 C, $CH_2$), 27.633 (1 C, $CH_2$), 28.225 (6 C, $CH_3$), 35.680 (1 C, $CH_2$), 40.061 (1 C, $CH_2$), 44.437 (1 C, $CH_2$), 46.713 (1 C, $CH_2$), 78.872 (1 C, quat. C $^t$Bu), 79.380 (1 C, quat. C $^t$Bu), 123.180 (2 C, Ar-CH), 132.054 (2 C, Ar-C), 133.941 (2 C, Ar-CH), 155.430 (1 C, COO), 156.002 (1 C, COO), 168.303 (2 C, CON).
MS-ESI(+): $m/z$ = 499 ($M^+$ + Na), 514 ($M^+$ + K);
IR (ATR): ν = 3367.35, 2974.83, 2932.30, 1772.13, 1707.48, 1687.80, 1394.36, 1364.25, 1247.15, 1166.74, 1029.52, 719.42 $cm^{-1}$.

*Synthesis of 2-(3-(4-aminobutylamino)propyl)isoindoline-1,3-dione as TFA salt (**13**)*:
To a solution of **14** (1.00 g, 2.10 mmol) in 2 mL dichloromethane 2 mL of TFA was added to cleave the Boc protection groups. The reaction mixture was stirred for 4 h at RT. After evaporation of the solvent, the residue was treated with diethyl ether (20 mL, 2 times) and the resulting white solid was filtered off and washed with diethyl ether. The solid product was dried under vacuum to yielding **13** as solid, yield quantitative; $R_f$ = 0.37 (dichloromethane:methanol:triethylamine) (7.5:2:0.5) mixture.
$^1$H NMR (300 MHz, $CDCl_3$, 25 °C): δ = 1.54-1.61 (m, 4H, 2 x $CH_2$), 1.93 (quintuplet, $J$ = 6.8 Hz, 2H, $CH_2$), 2.80 (sextet, $J$ = 6.2 Hz, 2H, $CH_2NH_3^+$, converts to triplet on $D_2O$ exchange), 2.90 (quintuplet, $J$ = 5.6 Hz, 2H, $CH_2NH_2^+$, Converts to triplet on $D_2O$ exchange), 2.96 (quintuplet, $J$ = 5.2 Hz, 2H, $CH_2NH_2^+$, Converts to triplet on $D_2O$ exchange), 3.64 (t, $J$ = 6.8 Hz, 2H, $CH_2$), 7.84-7.90 (m, 4H of ArH and 3H of $NH_3^+$, which exchanges with $D_2O$), 8.64 (bs, 2H, $NH_2^+$, which exchanges with $D_2O$) ppm.

$^{13}C$ NMR (75 MHz, $CDCl_3$, 25 °C): δ = 22.928 (1 C, $CH_2$), 24.276 (1 C, $CH_2$), 25.371 (1 C, $CH_2$), 35.157 (1C, $CH_2$), 38.509 (1 C, $CH_2$), 45.042 (1 C, $CH_2$), 46.566 (1 C, $CH_2$), 123.724 (2 C, Ar-CH), 132.069 (2 C, Ar-C), 135.183 (2 C, Ar-CH), 168.890 (2 C, CON).
MS-ESI(+): *m/z* = 276 [$M^+$ +1 (free amine)].
IR (ATR): ν = 2946.01, 2854.43, 1710.04, 1667.20, 1616.91, 1431.88, 1395.45, 1361.97, 1195.78, 1178.34, 1126.70, 831.09, 796.37, 718.50 $cm^{-1}$.

*Synthesis of tert-butyl 2,2'-(4-((2-tert-butoxy-2-oxoethyl)(3-(1,3-dioxoisoindolin-2-yl)propyl)amino) butylazane-diyl)diacetate (**12**)*:
To a solution of **13** (2.5 g, 5.0 mmol) in 1,4-dioxan (65 mL), *tert*-butyl bromoacetate (3.4 g, 2.82 mL, 17.4 mmol) and diisopropylethyl amine (DIEA) (3.2 g, 30.0 mmol) was added. The reaction mixture was stirred for 24h at 60°C. After completion of the reaction (tlc), the reaction mixture was filtered and filtrate was evaporated under vacuum. The residue was purified by column chromatography ($SiO_2$, dichloromethane:methanol) yielding **12** as a light greenish liquid, yield 1.9 g (3.0 mmol, 62.0 %); $R_f$ = 0.75 dichloromethane:methanol (97:3).
$^1H$ NMR (300 MHz, $CDCl_3$, 25 °C): δ = 1.42 (s, 9H, 3 x $CH_3$), 1.44 (bs, 18H, 6 x $CH_3$ and 2H of $CH_2$ merge under the singlet), 1.73 (broad quintuplet, 2H, $CH_2$), 1.80 (quintuplet, $J$ = 7.2 Hz, 2H, $CH_2$), 2.57 (t, $J$ = 7.0 Hz, 2H, $CH_2$), 2.66 (t, $J$ = 7.0 Hz, 4H, 2 x $CH_2$), 3.22 (s, 2H, $CH_2$), 3.40 (s, 4H, 2 x $NCH_2$), 3.72 (t, $J$ = 7.2 Hz, 2H, $CH_2$), 7.69 (dd, $J_1$ = 2.8 Hz, $J_2$ = 5.2 Hz, 2H, ArH), 7.82 (dd, $J_1$ = 2.8 Hz, $J_2$ = 5.2 Hz, 2H, ArH) ppm.
$^{13}C$ NMR (75 MHz, $CDCl_3$, 25 °C): δ = 25.117 (1C, $CH_2$), 25.591 (1 C, $CH_2$), 26.527 (1 C, $CH_2$), 28.018 (9 C, $CH_3$), 36.144 (1 C, $CH_2$), 51.550 (1 C, $CH_2$), 53.662 (1 C, $CH_2$), 53.952 (1 C, $CH_2$), 55.197 (1 C, $CH_2$), 55.738 (2 C, $CH_2$), 80.591 (1 C, quat. C $^t$Bu), 80.684 (2 C, quat. C $^t$Bu), 123.083 (2 C, Ar-CH), 132.196 (2 C, Ar-C), 133.797 (2 C, Ar-CH), 168.367 (2 C, CON), 170.774 (1 C, COO), 170.844 (2 C, COO).
MS-ESI(+): *m/z* = 618 [$M^+$+H].
IR (ATR): ν = 2976.69, 2934.86, 1771.96, 1709.98, 1467.22, 1392.99, 1366.37, 1251.52, 1215.52, 1146.15, 1036.89, 843.25, 719.75 $cm^{-1}$.

*Synthesis of tert-butyl 2,2'-(4-((3-aminopropyl)(2-tert-butoxy-2-oxoethyl)amino)butylazanediyl) diacetate (**11**)*:
To a solution of **12** (2.0 g, 3.2 mmol) in ethanol (30 mL), hydrazine hydrate (0.486 mL, 9.7 mmol) was added and reaction mixture was stirred for 8 h at 50°C. The formed solid precipitate was removed by filtration. The filtrate was concentrated in vacuum and the residue was dissolved in dichloromethane and washed with 10% KOH solution. The organic layer was extracted with brine (50 mL), dried over $Na_2SO_4$ and concentrated in vacuum yielding a residue which was further purified by column chromatography ($SiO_2$, dichloromethane:methanol:triethylamine). **10** was isolated as transparent liquid, yield 1.29 g (2.65 mmol, 81.8 %); $R_f$ = 0.65 dichloromethane:methanol:triethylamine (9:0.5:0.5).
$^1H$ NMR (300 MHz, $CDCl_3$, 25 °C): δ = 1.43 (s, 9H, 3 x $CH_3$), 1.44 (s, 18H, 6 x $CH_3$), 1.43-1.48 (m, 4H, 2 x $CH_2$ signals merge under methyl peaks), 1.57 (quintuplet, $J$ = 6.9 Hz, 2H, $CH_2$), 2.54 (t, $J$ = 6.8 Hz, 2H, $CH_2$), 2.59 (t, $J$ = 7.2 Hz, 2H, $CH_2$), 2.67 (t, $J$ = 7.0 Hz, 2H, $CH_2$), 2.72 (t, $J$ = 6.6 Hz, 2H, $CH_2$), 3.20 (s, 2H, $NCH_2$), 3.4 (s, 4H, 2 x $NCH_2$) ppm.
$^{13}C$ NMR (75 MHz, $CDCl_3$, 25 °C): δ = 24.904 (1 C, $CH_2$), 25.658 (1 C, $CH_2$), 28.025 (9 C, $CH_3$), 31.088 (1 C, $CH_2$), 40.361 (1 C, $CH_2$), 51.777 (1 C, $CH_2$), 53.913 (1 C, $CH_2$), 53.972 (1 C, $CH_2$), 55.772 (1 C, $CH_2$), 55.820 (2 C, $CH_2$), 80.581 (1 C, quat. C $^t$Bu), 80.719 (2 C, quat. C $^t$Bu), 170.772 (1 C, COO), 170.985 (2 C, COO) ppm.
MS-ESI(+): *m/z* = 488 [$M^+$+H].
IR (ATR): ν = 2977.56, 2852.76, 1730.27, 1456.53, 1392.13, 1367.09, 1254.25, 1218.38, 1146.44, 841.75, 733.85, 701.66 $cm^{-1}$.

**Synthesis of di-tert-butyl 2,2'-((3-((4-((3-aminopropyl)(2-(tert-butoxy)-2-oxoethyl)amino)butyl)(2-(tert-butoxy)-2-oxoethyl)amino)propyl)azanediyl)diacetate (17)**

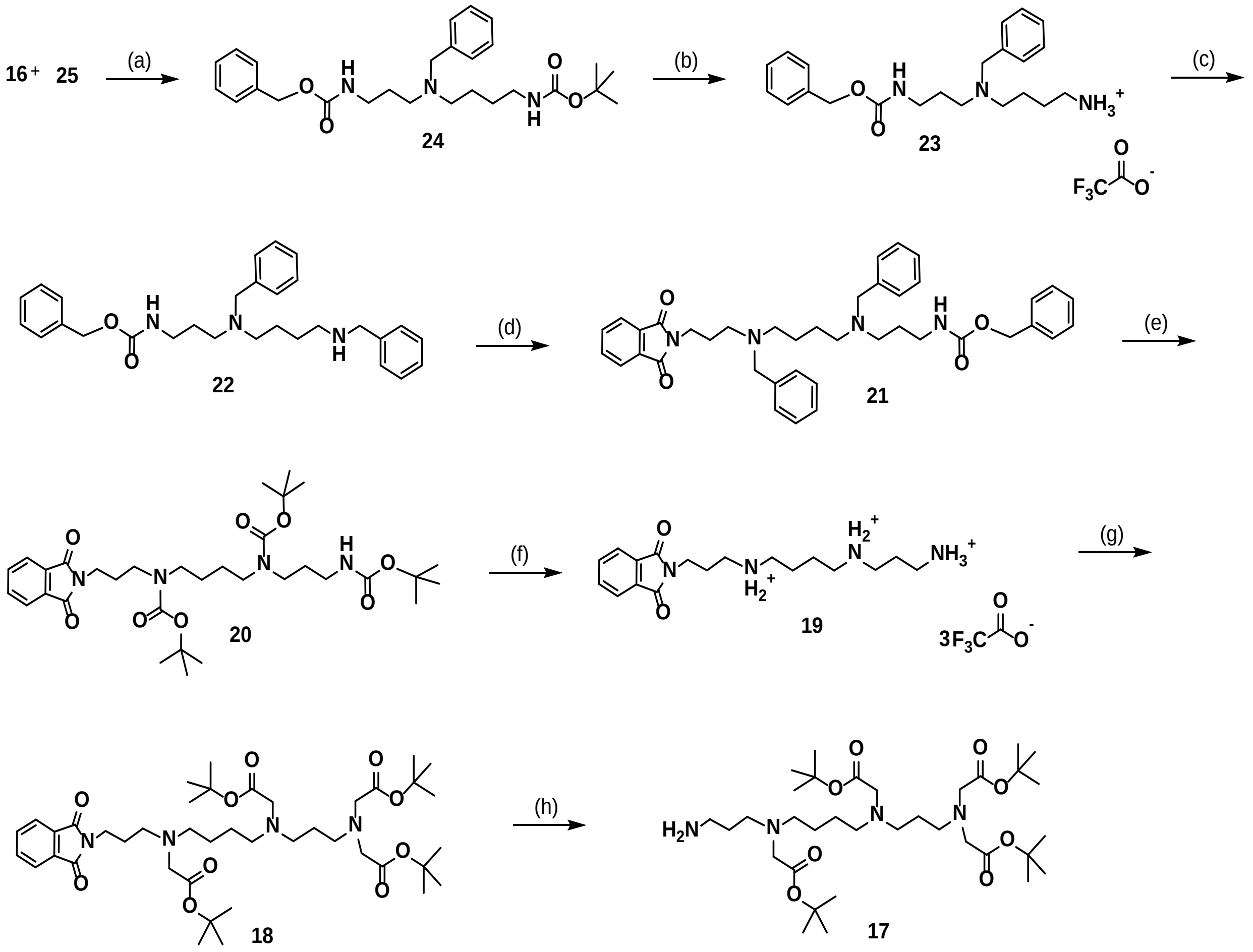

**Scheme 3.** Synthesis of precursor amine **17**. (a) **16**, *N*-carbobenzyloxy-3-bromopropylamine (**25**), chloroform:acetonitrile (1:1), $Na_2CO_3$, 70°C, 48 h; (b) trifluoroacetic acid, $CH_2Cl_2$, 4 h, RT; (c) benzaldehyde, $MgSO_4$, triethylamine, $CH_3OH$, overnight, then $NaBH_4$ addition at 0°C for 1 h and then RT for 1 h; (d) *N*-(3-bromopropyl)phthalimide, chloroform:acetonitrile (1:1), $Na_2CO_3$, 80°C, 48 h; (e) 10% Pd/C, $H_2$, di-*tert*-butyl dicarbonate, acetic acid (cat.), $CH_3OH$; (f) trifluoroacetic acid, $CH_2Cl_2$, 4 h, RT; (g) *tert*-butyl 2-bromoacetate, 1,4-dioxan-acetonitrile (8:2), DIEA, 45°C, 24 h; (h) hydrazine hydrate, ethanol, 50°C, 6 h.

*Synthesis of tert-butyl 4-(benzyl(3-(carbobenzyloxyamino)propyl)amino)butylcarbamate (**12**)*:
To a solution of compound **16** (10.00 g, 36.0 mmol) in acetonitrile:chloroform (1:1) mixture (100 mL), *N*-carbobenzyloxy-3-bromopropylamine (11.29 g, 42.0 mmol) and $Na_2CO_3$ (3.80 g, 36.0 mmol) was added. The reaction mixture was stirred for 48 h at 70°C. After completion of the reaction (tlc), the reaction mixture was filtered and the solvent was concentrated in *vacuo*. The residue was purified through column chromatography ($SiO_2$, dichloromethane:methanol). **24** was isolated as a liquid, yield 12.3 g (26.3 mmol, 73.2 %); $R_f$ = 0.65 (dichloromethane:methanol, 95:5).
$^1$H NMR (300 MHz, $CDCl_3$, 25 °C): δ = 1.43 (s, 9H, 3 x $CH_3$); 1.45-1.50 (m, 4H, 2 x $CH_2$), 1.64 (quintuplet, *J* = 6.1 Hz, 2H, $CH_2$), 2.39 (t, *J* = 6.8 Hz, 2H, $CH_2$), 2.45 (t, *J* = 6.2 Hz, 2H, $CH_2$), 3.04 (bq, *J* = 5.6 Hz, 2H, $CH_2$), 3.20 (q, *J* = 6.0 Hz, 2H, $CH_2$), 3.49 (s, 2H, $NCH_2$), 4.61 (bt, 1H, NHCO), 5.08 (s, 2H, $OCH_2$), 5.77 (bt, 1H, NHCO), 7.20-7.36 (m, 10H, ArH-benzyl) ppm.

$^{13}C$ NMR (75 MHz, $CDCl_3$, 25°C): δ = 24.040 (1 C, $CH_2$), 26.268 (1 C, $CH_2$), 27.693 (1 C, $CH_2$), 28.263 (3 C, $CH_3$), 40.132 (1 C, $CH_2$), 40.204 (1 C, $CH_2$), 52.105 (1 C, $CH_2$), 53.252 (1 C, $CH_2$), 58.701 (1 C, $CH_2$), 66.207 (1 C, $CH_2$), 78.849 (1 C, quat. C $^t$Bu), 126.956 (1 C, Ar-CH), 127.904 (1 C, Ar-CH), 127.941 (2 C, Ar-CH), 128.232 (2 C, Ar-CH), 128.406 (2 C, Ar-CH), 128.907 (2 C, Ar-CH), 136.855 (1 C, Ar-C), 139.210 (1 C, Ar-C), 156.006 (1 C, CO(NH)O), 156.352 (1 C, CO(NH)O) ppm.
MS-ESI(+): *m*/*z* = 469 [$M^+$].
IR (ATR): ν = 3337.67, 2933.91, 2865.17, 2802.95, 1689.06, 1514.50, 1453.38, 1364.86, 1247.62, 1167.31, 733.76, 696.73 $cm^{-1}$.

*Synthesis of benzyl 3-((4-aminobutyl)(benzyl)amino)propylcarbamate as TFA salt (**23**)*:
5 mL of trifluoroacetic acid (TFA) was added to a solution of **24** (4.3 g, 9.2 mmol) in 5 mL dichloromethane. The reaction mixture was stirred for 4h at room temperature. After evaporation of the solvent, the residue was dispersed in diethyl ether to precipitate the product. The product was filtered and dried under vacuum. **23** was isolated as TFA salt, yield 2.9 g (6.0 mmol, 65.8 %); $R_f$ = 0.3 (dichloromethane:methanol 87:13).
$^1H$ NMR (300 MHz, $CDCl_3$, 25 °C): δ = 1.47-1.52 (m, 4H, 2 x $CH_2$), 1.62 (quintuplet, *J* = 6.2 Hz, 2H, $CH_2$), 2.41 (t, *J* = 6.6 Hz, 2H, $CH_2$), 2.44 (t, *J* = 6.3 Hz, 2H, $CH_2$), 2.75 (sextet, *J* = 6.0 Hz, 2H, $CH_2NH_3^+$ converts to triplet on $D_2O$ exchange), 3.18 (q, *J* = 6.1 Hz, 2H, $CH_2$), 3.50 (s, 2H, $NCH_2$), 5.04 (s, 2H, $OCH_2$), 5.76 (bt, 1H, NHCO), 7.19-7.35 (m, 10H, ArH-benzyl) ppm.
$^{13}C$ NMR (75 MHz, $CDCl_3$, 25 °C): δ = 20.233 (1 C, $CH_2$), 23.651 (1 C, $CH_2$), 24.054 (1 C, $CH_2$), 38.704 (1 C, $CH_2$), 45.688 (1 C, $CH_2$), 49.315 (1 C, $CH_2$), 51.555 (1 C, $CH_2$), 56.718 (1 C, $CH_2$), 66.514 (1 C, $CH_2$), 127.820 (1 C, Ar-CH), 128.077 (1 C, Ar-CH), 128.328 (1 C, Ar-C), 128.519 (2 C, Ar-CH), 129.388 (2 C, Ar-CH), 130.158 (2 C, Ar-CH), 130.878 (2 C, Ar-CH), 136.603 (1 C, Ar-C), 157.162 (1 C, CO(NH)O) ppm.
MS-ESI(+): *m*/*z* = 370 [$M^+$ + 1].
IR (ATR): ν = 3306.30, 2956.68, 1669.28, 1531.22, 1456.17, 1261.20, 1198.55, 1126.48, 1027.59, 833.37, 798.93, 741.42, 720.43, 698.93 $cm^{-1}$.

*Synthesis of benzyl 3-(benzyl(4-(benzylamino)butyl)amino)propylcarbamate (**22**)*:
Compound **23** (2.5 g, 6.8 mmol) was dissolved in 25 mL of methanol. To this solution, benzaldehyde (0.86 g, 8.1 mmol), $MgSO_4$ (1.63 g, 13.5 mmol) and triethylamine (0.81 g, 8.1 mmol) was added and the reaction mixture was stirred at RT overnight. After cooling to 0°C, $NaBH_4$ (1.53 g, 40.6 mmol) was added over a period of 1 h. The reaction mixture was further stirred at 0°C for 1 h and then at RT for an additional hour. Water (200 mL) was added to quench the reaction and the solution was extracted with ethyl acetate (3 x 150 mL). The combined organic layers were dried over $Na_2SO_4$ and afterwards, the solvent was concentrated in *vacuo*. The residue was purified through column chromatography ($SiO_2$, dichloromethane:methanol) to isolate pure **22** as a solid, yield 1.62 g (3.5 mmol, 68.3 %); $R_f$ = 0.55 (dichloromethane:methanol, 90:10).
$^1H$ NMR (300 MHz, $CDCl_3$, 25 °C): δ = 1.50 (broad quintuplet, 4H, 2 x $CH_2$), 1.61 (quintuplet, *J* = 6.2 Hz, 2H, $CH_2$), 2.37 (bt, 2H, $CH_2$), 2.43 (t, *J* = 6.2 Hz, 2H, $CH_2$), 2.57 (bt, 2H, $CH_2$), 3.18 (q, *J* = 5.7 Hz, 2H, $CH_2$), 3.48 (s, 2H, $CH_2N$), 3.74 (s, 2H, $CH_2N$), 5.05 (s, 2H, $CH_2O$), 5.78 (bt, 1H, NHCO), 7.19-7.33 (m, 15H, ArH-benzyl) ppm.
$^{13}C$ NMR (75 MHz, $CDCl_3$, 25 °C): δ = 24.520 (2 C, $CH_2$), 26.218 (1 C, $CH_2$), 27.288 (1 C, $CH_2$), 40.169 (1 C, $CH_2$), 48.731 (1 C, $CH_2$), 52.031 (1 C, $CH_2$), 53.496 (1 C, $CH_2$), 58.674 (1 C, $CH_2$), 66.258 (1 C, $CH_2$), 127.029 (2 C, Ar-CH), 127.214 (1 C, Ar-CH), 127.940 (2 C, Ar-CH), 128.303 (2 C, Ar-CH), 128.376 (2 C, Ar-CH), 128.455 (2 C, Ar-CH), 128.469 (2 C, Ar-CH), 129.001 (2 C, Ar-CH), 136.924 (2 C, Ar-C), 139.201 (1 C, Ar-C), 156.440 (1 C, CO(NH)O) ppm.
MS-ESI(+): *m*/*z* = 460 [$M^+$ +1].
IR (ATR): ν = 3327.82, 3061.76, 3028.48, 2935.02, 2801.81, 1702.25, 1513.74, 1495.03, 1452.96, 1248.01, 1129.53, 1026.97, 732.80, 695.96 $cm^{-1}$.

*Synthesis of benzyl 3-(benzyl(4-(benzyl(3-(1,3-dioxoisoindolin-2-yl)propyl)amino)butyl)amino) propylcarbamate (**21**)*:

To a solution of compound **22** (6.0 g, 13.0 mmol) in acetonitrile:chloroform (1:1) (150 mL), N-(3-bromopropyl)phthalimide (4.2 g, 16.0 mmol) and $Na_2CO_3$ (3.8 g, 13.0 mmol) was added. The reaction mixture was stirred for 3 days at 75°C. After completion of the reaction (tlc), the reaction mixture was filtered and the solvent was concentrated in *vacuo*. The residue was purified by column chromatography ($SiO_2$, dichloromethane:methanol) yielding **21** as a liquid, yield 4.83 g (7.4 mmol, 57.3 %); $R_f$ = 0.75 (dichloromethane:methanol, 95:5).

$^1$H NMR (300 MHz, $CDCl_3$, 25 °C): δ = 1.40-1.47 (m, 4H, 2 x $CH_2$); 1.59-1.62 (m, 2H, $CH_2$); 1.78 (quintuplet, $J$ = 7.1 Hz, 2H, $CH_2$), 2.31-2.37 (m, 4H, 2 x $CH_2$); 2.41-2.45 (m, 4H, 2 x $CH_2$); 3.19 (q, $J$ = 5.7 Hz, 2H, $CH_2$), 3.47 (s, 2H, $CH_2N$), 3.48 (s, 2H, $CH_2N$), 3.65 (t, $J$ = 7.6 Hz, 2H, $CH_2$), 5.04 (s, 2H, $CH_2O$), 5.84 (bt, 1H, NHCO), 7.15-7.34 (m, 15H, ArH-benzyl); 7.66 (dd, $J_1$ = 3.2 Hz, $J_2$ = 5.6 Hz, 2H, ArH), 7.80 (dd, $J_1$ = 2.8 Hz, $J_2$ = 5.2 Hz, 2H, ArH) ppm.

$^{13}$C NMR (75 MHz, $CDCl_3$, 25 °C): δ = 24.441 (1 C, $CH_2$), 24.689 (1 C, $CH_2$), 25.958 (1 C, $CH_2$), 26.224 (1 C, $CH_2$), 36.302 (1 C, $CH_2$), 40.240 (1 C, $CH_2$), 50.999 (1 C, $CH_2$), 52.101 (1 C, $CH_2$), 53.344 (1 C, $CH_2$), 53.602 (1 C, $CH_2$), 58.419 (1 C, $CH_2$), 58.725 (1 C, $CH_2$), 66.178 (1 C, $CH_2$), 123.094 (2 C, Ar-CH), 126.707 (1 C, Ar-CH), 126.885 (1 C, Ar-CH), 127.872 (1 C, Ar-CH), 127.911 (2 C, Ar-CH), 128.095 (2 C, Ar-CH), 128.224 (2 C, Ar-CH), 128.410 (2 C, Ar-CH), 128.810 (2 C, Ar-CH), 128.934 (1 C, Ar-CH), 132.164 (1 C, Ar-CH), 133.803 (2 C, Ar-CH), 136.953 (1 C, Ar-C), 139.450 (2 C, Ar-C), 139.752 (2 C, Ar-C), 156.404 (1 C, CO(NH)O), 168.405 (2 C, CON) ppm.

MS-ESI(+): $m/z$ = 647 [$M^+$ + 1].

IR (ATR): ν = 3337.34, 3065.25, 3031.38, 2950.78, 1703.45, 1515.73, 1497.86, 1392.65, 1367.12, 1175.31, 741,54, 732.48, 721.59, 695.02 $cm^{-1}$.

*Synthesis of tert-butyl 3-((4-((3-(1,3-dioxoisoindolin-2-yl)propyl)(tert-butoxycarbonyl)amino) butyl) (tert-butoxycarbonyl)amino)propylcarbamate (**20**)*:

To the $N_2$-flushed solution of compound **21** (13.0 g, 20.0 mmol) in 200 mL of methanol (dry), 10% Pd/C (2.0 gm) was added followed by addition of actic acid (4.0 mL). Afterwards, di-*tert*-butyldicarbonate (21.94 g, 100.0 mmol) was added to this solution. The flask was degassed and saturated with hydrogen and stirred for 48 h at room temperature. The degassing and saturation with hydrogen were regularly repeated during this time interval. The reaction progress was monitored by TLC. After completion of the reaction (tlc), the Pd/C was filtered off (celite), washed with methanol and the solvent was concentrated in *vacuo*. The residue was purified by column chromatography ($SiO_2$, dichloromethane:methanol) to isolate **20** as a liquid, yield 7.2 g (11.4 mmol, 56.6 %); $R_f$ = 0.65 (dichloromethane:methanol, 97:3).

$^1$H NMR (300 MHz, $CDCl_3$, 25 °C): δ = 1.39-1.47 [m which shows two large singlets at 1.42 (s, 18H, 6 x $CH_3$) and 1.43 (s, 9H, 3 x $CH_3$) along with 4H protons 2 x $CH_2$), 1.63 (bq, 2H, $CH_2$), 1.89 (quintuplet, $J$ = 5.4 Hz, 2H, $CH_2$), 3.08-3.19 (m, 10H, 5 x $CH_2$), 3.67 (t, $J$ = 5.1 Hz, 2H, $CH_2$), 5.35 (bs, 1H, NH of carbamate), 7.69-7.71 (m, 2H, ArH), 7.82-7.84 (m, 2H, ArH) ppm.

$^{13}$C NMR (75 MHz, $CDCl_3$, 25 °C): δ = 24.841 (1 C, $CH_2$), 25.189 (1 C, $CH_2$), 26.158 (1 C, $CH_2$, 26.824 (1 C, $CH_2$), 28.018 (6 C, $CH_3$), 28.832 (3 C, $CH_3$), 37.002 (1 C, $CH_2$), 40.840 (1 C, $CH_2$), 51.349 (1 C, $CH_2$), 52.631 (1 C, $CH_2$), 53.563 (1 C, $CH_2$), 53.832 (1 C, $CH_2$), 78.932 (1 C, quat. C $^t$Bu), 79.245 (2 C, quat. C $^t$Bu), 123.344 (2 C, Ar-CH), 132.094 (2 C, Ar-C), 133.911 (2 C, Ar-CH), 155.635 (1 C, COO), 156.052 (2 C, COO), 168.543 (2 C, CON) ppm.

MS-ESI(+): $m/z$ = 633 [$M^+$], 533 [$M^+$ - Boc].

IR (ATR): ν = 3357.45, 2976.63, 2942.34, 1778.33, 1709.58, 1690.12, 1398.65, 1369.05, 1251.75, 1169.23, 1035.52, 720.12 $cm^{-1}$.

*Synthesis of 2-(3-(4-(3-aminopropylamino)butylamino)propyl)isoindoline-1,3-dione as TFA salt (**19**)***:**

In order to cleave the Boc protecting groups, 4.5 mL of trifluoroacetic acid was added to a solution of **20** (1.50 g, 2.4 mmol) in 4.5 mL of dichloromethane. The reaction mixture was stirred for 4 h at RT. After evaporation of the solvent, the product was precipitated by addition of diethyl ether. After filtration, the solid product was dried under vacuum yielding **19**, as a white solid, yield quantitaive; $R_f$ = spot remained on base.
$^1$H NMR (300 MHz, $CDCl_3$, 25 °C): δ = 1.58 (broad quintuplet, 4H, 2 x $CH_2$), 1.83-1.95 (m, 4H, 2 x $CH_2$), 2.83-2.96 (m, 10H, 5 x $CH_2$), 3.63 (t, *J* = 6.6 Hz, 2H, $CH_2$), 7.82-7.88 (m, 4H, ArH) ppm.
$^{13}$C NMR (75 MHz, $CDCl_3$, 25 °C): δ = 22.748 (1 C, $CH_2$), 22.816 (1 C, $CH_2$), 23.779 (1 C, $CH_2$), 25.161 (1 C, $CH_2$), 34.972 (1 C, $CH_2$), 36.197 (1 C, $CH_2$), 44.015 (1 C, $CH_2$), 44.727 (1 C, $CH_2$), 46.210 (1 C, $CH_2$), 46.264 (1 C, $CH_2$), 123.366 (2 C, Ar-CH), 131.826 (2 C, Ar-C), 134.780 (2 C, Ar-CH), 168.385 (2 C, CON) ppm.
MS-ESI(+): *m*/*z* = 333 [$M^+$ +1 (free amine)].
IR (ATR): ν = 3036.21, 2946.64, 2857.78, 2554.55, 2494.87, 1776.31, 1711.60, 1665.91, 1612.97, 1431.96, 1396.14, 1362.30, 1196.51, 1175.84, 1126.59, 797.46, 773.51, 719.60 $cm^{-1}$.

*Synthesis of tert-butyl 2,2'-(3-((2-tert-butoxy-2-oxoethyl)(4-((2-tert-butoxy-2-oxoethyl)(3-(1,3-dioxoisoindolin-2-yl) propyl)amino)butyl)amino)propylazanediyl)diacetate (**18**)*:
To a solution of compound **19** (1.40 g, 2.0 mmol) in dioxane:acetonitrile (8:2) (20 mL), *tert*-butyl bromoacetate (1.5 mL, 9.1 mmol) and diisopropylethylamine (2.90 mL, 16.5 mmol) were added. The reaction mixture was stirred for 24h at 45°C. After completion of the reaction (tlc), the reaction mixture was filtered and the solvent was concentrated in *vacuo*. The residue was purified by column chromatography ($SiO_2$, dichloromethane:methanol) yielding **18** as a liquid, yield 0.33 g (0.42 mmol, 20.1 %); $R_f$ = 0.8 dichloromethane: methanol, (97:3).
$^1$H NMR (300 MHz, $CDCl_3$, 25 °C): δ = 1.42 (s, 18H, 6 x $CH_3$), 1.43 (s, 18H, 6 x $CH_3$ These two singlets superimpose on the signals of 4H protons, 2 x $CH_2$), 1.61 (quintuplet, *J* = 7.3 Hz, 2H, $CH_2$), 1.80 (quintuplet, *J* = 7.1 Hz, 2H, $CH_2$), 2.56-2.71 (m, 10H, 5 x $CH_2$), 3.19 (s, 2H, $NCH_2$), 3.22 (s, 2H, $NCH_2$), 3.40 (s, 4H, 2 x $NCH_2$), 3.72 (t, *J* = 7.4 Hz, 2H, $CH_2$), 7.69 (dd, $J_1$ = 3.2 Hz, $J_2$ = 5.6 Hz, 2H, ArH), 7.82 (dd, $J_1$ = 3.2 Hz, $J_2$ = 5.6 Hz, 2H, ArH) ppm.
$^{13}$C NMR (75 MHz, $CDCl_3$, 25 °C): δ = 25.186 (1 C, $CH_2$), 25.314 (1 C, $CH_2$), 26.062 (1 C, $CH_2$), 26.529 (1 C, $CH_2$), 28.019 (12 C, $CH_3$), 36.146 (1 C, $CH_2$), 51.570 (1 C, $CH_2$), 51.872 (1 C, $CH_2$), 52.091 (1 C, $CH_2$), 53.739 (1 C, $CH_2$), 54.138 (1 C, $CH_2$), 55.231 (1 C, $CH_2$), 55.531 (1 C, $CH_2$), 55.712, (2 C, $CH_2$), 80.466 (1 C, quat. C $^t$Bu), 80.566 (1 C, quat. C $^t$Bu), 80.664 (2 C, quat. C $^t$Bu), 123.076 (2 C, Ar-CH), 132.200 (2 C, Ar-C), 133.786 (2 C, Ar-CH), 168.349 (2 C, CON), 170.728 (1 C, COO), 170.823 (1 C, COO), 170.908 (2 C, COO) ppm.
MS-ESI(+): *m*/*z* = 789 [$M^+$ +1].
IR (ATR): ν = 2977.48, 2944.89, 1711.63, 1392.94, 1366.75, 1254.09, 1216.91, 1146.28, 1037.53, 844.16, 734.94, 721.37, 529.90 $cm^{-1}$.

*Synthesis of tert-butyl 2,2'-(3-((4-((3-aminopropyl)(2-tert-butoxy-2-oxoethyl)amino)butyl)(2-tert-butoxy-2-oxoethyl)amino)propylazanediyl)diacetate (**17**)*:
To a solution of compound **18** (0.40 g, 0.5 mmol) in ethanol (20 mL), hydrazine hydrate (0.076 mL, 1.5 mmol) was added and the reaction mixture was stirred for 4 h at 45°C. The formed precipitate was filtered off and the filtrate was concentrated in vacuum. The residue was dissolved in ethyl acetate and washed with 10% KOH solution. The organic layer was extracted with brine (50 mL), dried over $Na_2SO_4$ and concentrated in vacuum to obtained a residue which was further purified by column chromatography ($SiO_2$, dichloromethane:methanol:triethylamine) yielding **17** as a liquid, yield 0.18 g (0.28 mmol, 55.5 %); $R_f$ = 0.45 (dichloromethane:methanol:triethylamine, 9.5:0.25:0.25).
$^1$H NMR (300 MHz, $CDCl_3$, 25 °C): δ = 1.43 (s, 18H, 6 x $CH_3$), 1.44 (s, 18H, 6 x $CH_3$ These two singlets superimpose on the signals of 4H protons 2 x $CH_2$), 1.57-1.66 (m, 4H, 2 x $CH_2$), 2.46 (bs, 2H,

$NH_2$), 2.53-2.61 (m, 8H, 4 x $CH_2$), 2.69 (t, *J* = 7.4 Hz, 2H, $CH_2$), 2.79 (t, *J* = 6.4 Hz, 2H, $CH_2$), 3.18 (s, 2H, $NCH_2$), 3.19 (s, 2H, $NCH_2$), 3.41 (s, 4H, $NCH_2$) ppm.
$^{13}C$ NMR (75 MHz, $CDCl_3$, 25 °C): δ = 24.877 (1 C, $CH_2$), 25.239 (1 C, $CH_2$), 26.021 (1 C, $CH_2$), 28.007 (12 C, $CH_3$), 29.730 (1 C, $CH_2$), 40.240 (1 C, $CH_2$), 51.879 (1 C, $CH_2$), 52.003 (1 C, $CH_2$), 52.083 (1 C, $CH_2$), 53.962 (1 C, $CH_2$), 54.074 (1 C, $CH_2$), 55.668 (1 C, $CH_2$), 55.730 (1 C, $CH_2$), 55.862 (2 C, $CH_2$), 80.524 (1 C, quat. C $^t$Bu), 80.724 (1 C, quat. C $^t$Bu), 80.785 (2 C, quat. C $^t$Bu), 170.738 (1 C, COO), 170.929 (1 C, COO), 171.076 (2 C, COO) ppm.
MS-ESI(+): *m/z* = 659 [$M^+$ +1].
IR (ATR): ν = 2975.56, 2934.63, 1727.65, 1456.90, 1391.89, 1366.63, 1251.94, 1130.35, 1246.41, 938.13, 843.93, 749.82, 579.35, 457.19, 433.55 $cm^{-1}$.

## Solubility tests for the characterization of (1a) in solution

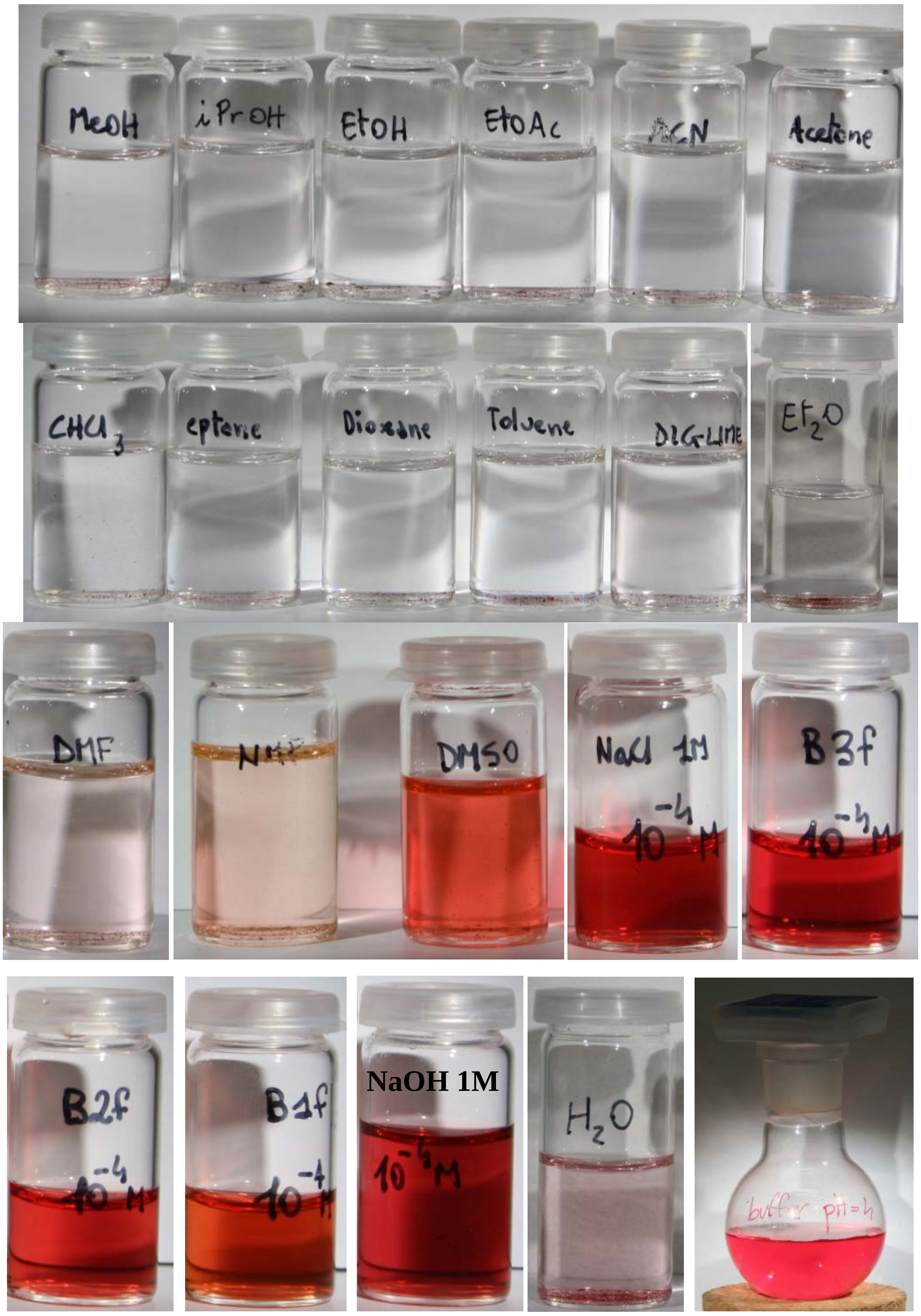

Figure S1. Solubility test for **1a.** Legend: B1f → buffer pH = 10 (VWR, $H_3BO_3/KCl/NaOH$); B2f → buffer pH = 10 (Fisher, $K_2CO_3/K_2B_8O_{13}/KOH$); B3f → buffer pH = 7 (VWR, $KH_2PO_4/Na_2HPO_4$); Buffer pH = 4 (Fisher, $C_8H_5KO_4$).

## Water-induced aggregation of (1a) in DMSO solution

The [1]H-NMR spectrum of **1a** in DMSO-$d_6$ was recorded in presence of TFA and HCl 37% (0.05 mL). As shown in Figure S2, all peaks in the aliphatic region ($\delta_H$ 1.6 – 4.2 ppm) remain more or less at the same position in both samples, while the aromatic area ($\delta_H$ 8.2 – 8.9 ppm) changes completely. In the presence of TFA, two well resolved doublets are visible and each of them displays a broad undefined band on the side.

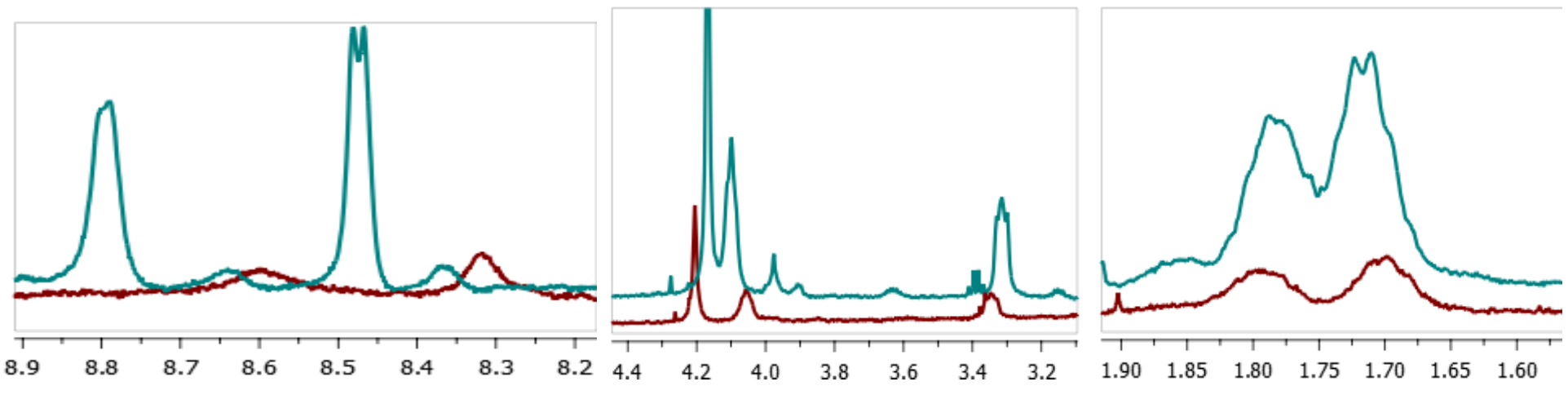

Figure S2. $^1$H-NMR spectrum of **1a** in DMSO-$d_6$ with addition of 1 drop of TFA (blue) or HCl (red)

When HCl is added, instead of TFA, just those two broad bands are visible, whereas the two doublets are lost. Since the addition of concentrated HCl adds some water to the sample, it is assumed that the PBI derivative **1a** aggregates more strongly under these conditions. Therefore, the two broad bands are attributed to the aggregated form of **1a**. To further support these preliminary observations, a titration experiment followed by UV/Vis measurements (Figure S3) was carried out.

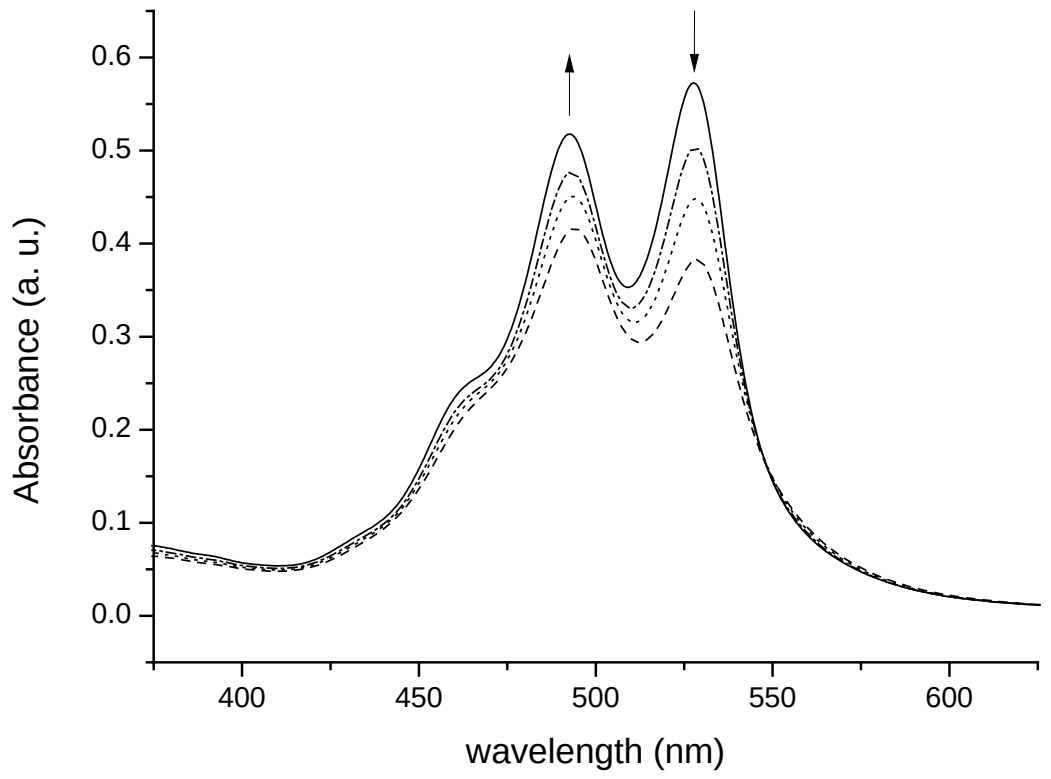

Figure S3. Absorption spectra of PBI **1a** in DMSO upon addition of conc. HCl
(arrows show spectrum modifications upon acid addition)

A solution of PBI **1a** in DMSO was titrated with concentrated HCl (37 %) and the UV/Vis spectra were collected after each addition. As presented in Figure 4, upon addition of HCl the spectrum profile changes significantly: the intensity of the (0,0) peak decreases and at the same time an increase of the intensity of the (0,1) peak is observed. This suggests that aggregation of the perylene bisimide surfactant dye **1a** is taking place.

## Complexation experiments for (1a) in water solution

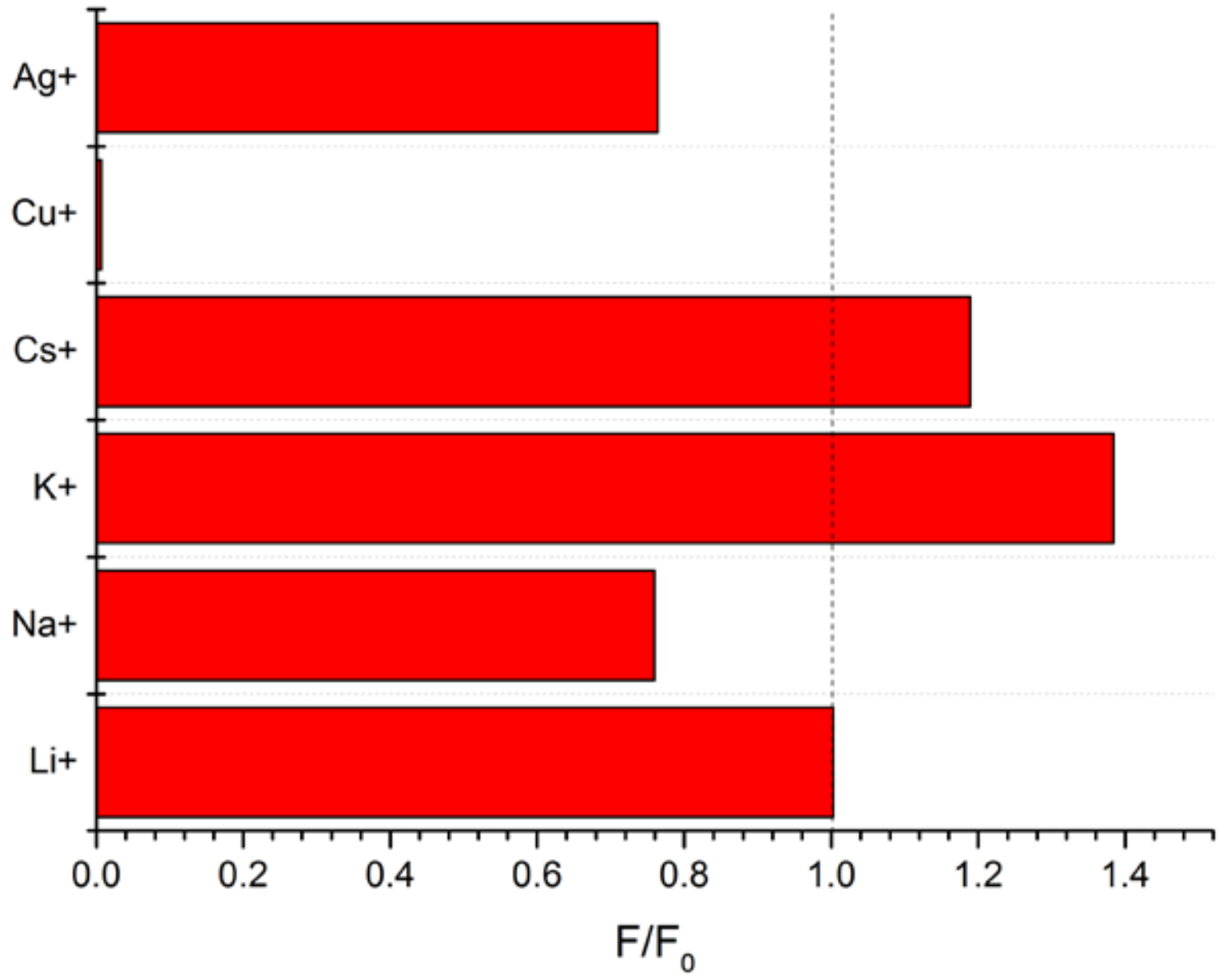

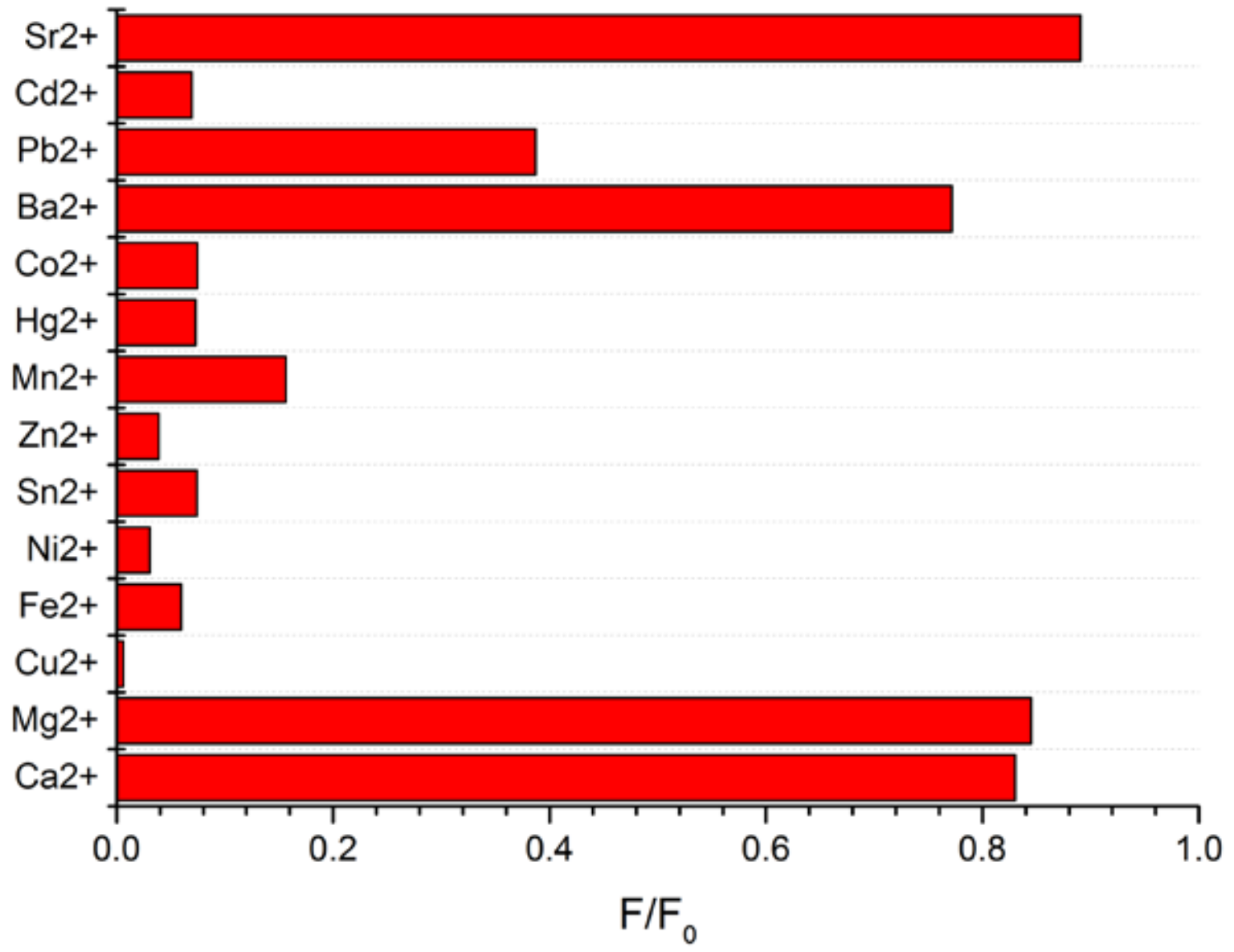

Figure S4. $F/F_0$ ratio for mono- (top), di- (bottom) valent ions – **1a** complexes in water

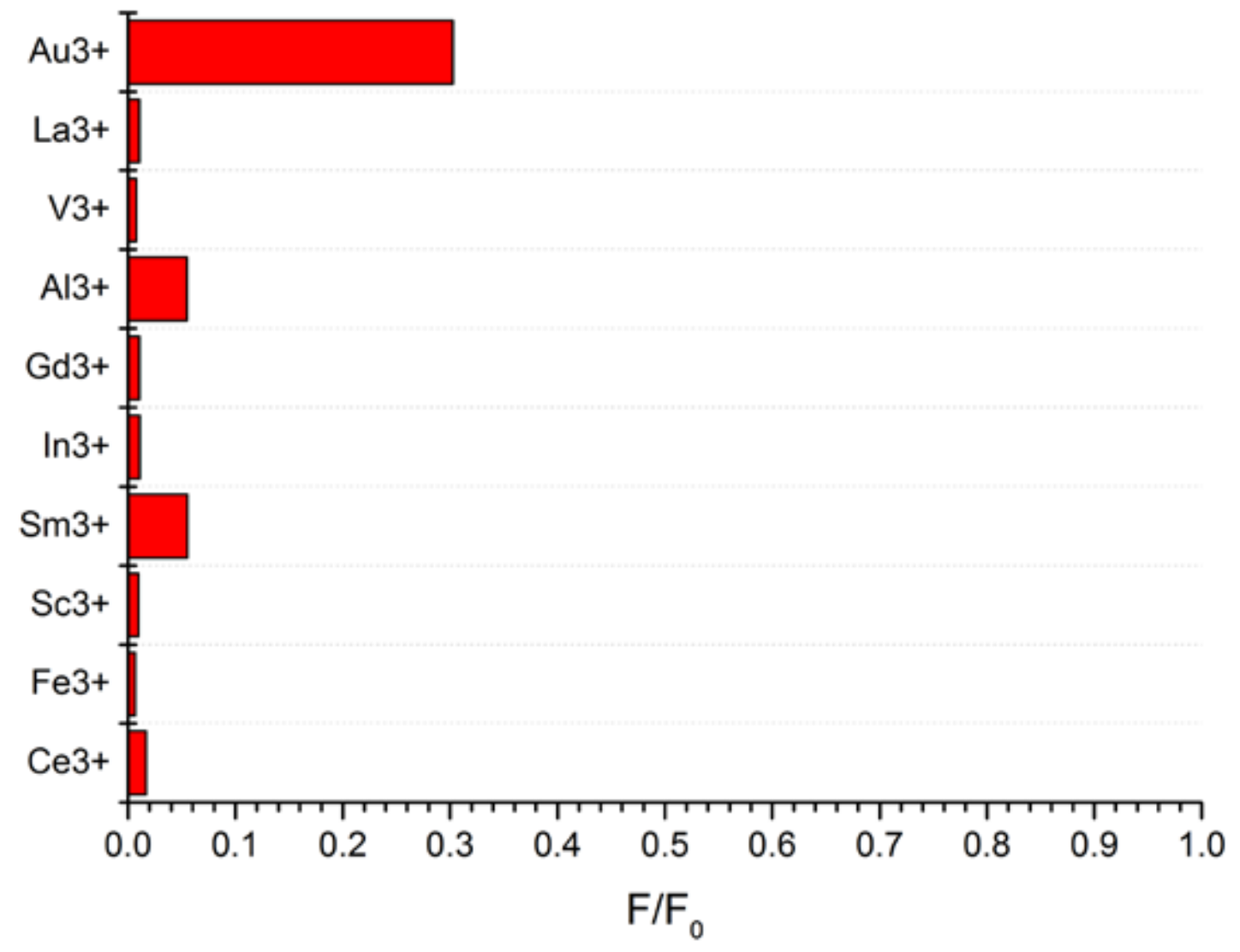

Figure S5. $F/F_0$ ratio for trivalent ions – **1a** complexes in water

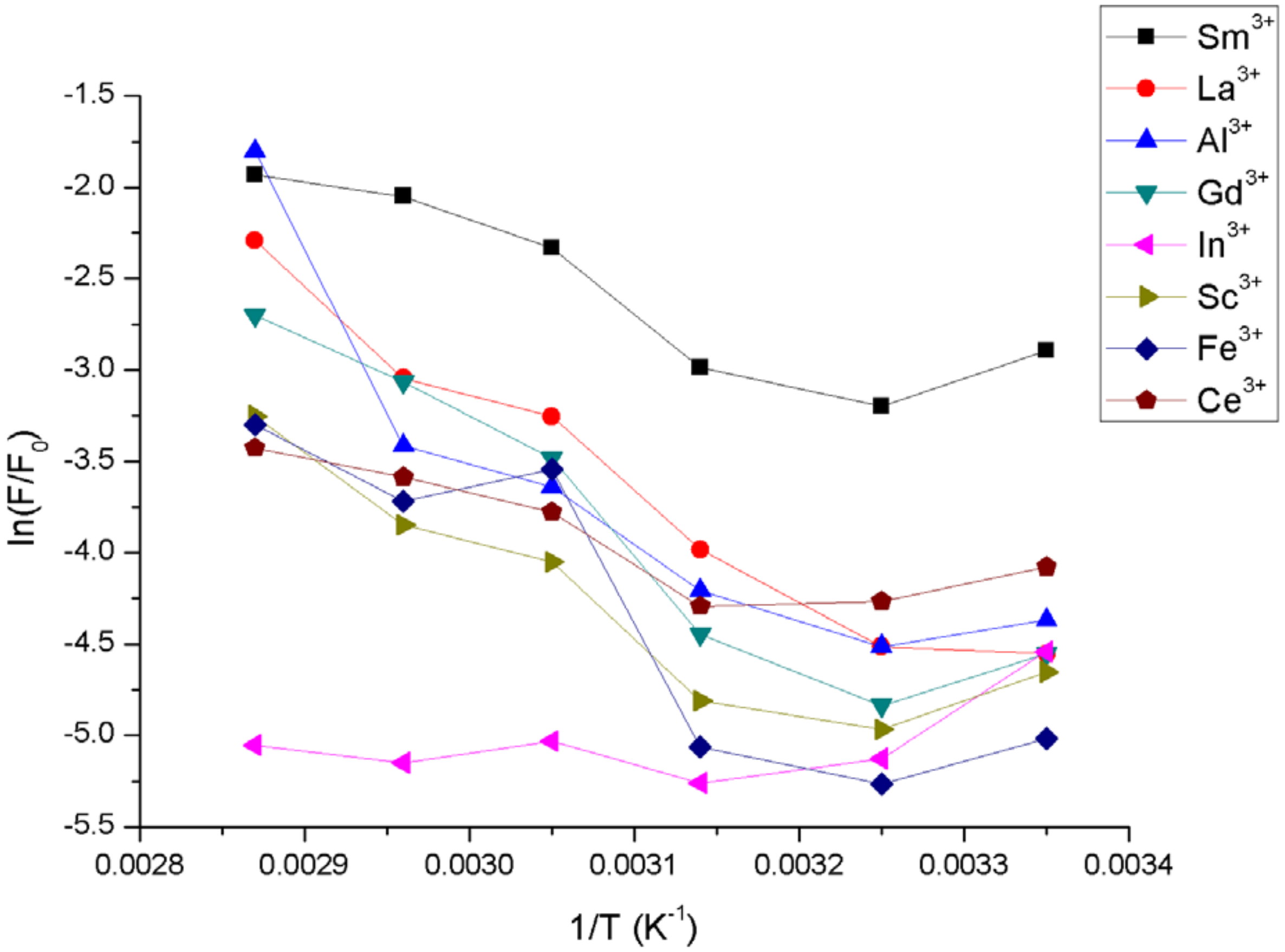

Figure S6. Van't Hoff plots for the stability trivalent ions – **1a** complexes in water by changing the temperature